\documentclass[11pt]{article}
\usepackage{times}
\usepackage{fullpage}
\usepackage{latexsym}
\usepackage{amsmath}
\usepackage{psfig}

\newtheorem{theorem}{Theorem}[section]
\newtheorem{lemma}[theorem]{Lemma}

\newtheorem{problem}[theorem]{Problem}

\newcommand{\qed}{\hfill $\Box$}
\newenvironment{proof}{{\bf Proof.~}}{\qed}

\newcommand{\ignore}[1]{}

\newcommand{\CA}{\ensuremath{{\cal A}}}
\newcommand{\CC}{\ensuremath{{\cal C}}}
\newcommand{\CD}{\ensuremath{{\cal D}}}
\newcommand{\CH}{\ensuremath{{\cal H}}}
\newcommand{\CP}{\ensuremath{{\cal P}}}
\newcommand{\CS}{\ensuremath{{\cal S}}}
\newcommand{\HC}{\ensuremath{H_{\CC}}}
\newcommand{\equ}[1]{(\ref{#1})}

\newcommand{\figinbox}[2]{\parbox{#1}{\psfig{file=#2,width=#1}}}
\newcommand{\pref}{\text{\it pref}}
\newcommand{\opt}{\text{\it opt}}
\newcommand{\cost}{\text{\it cost}}

\title{Improving Table Compression with Combinatorial Optimization\thanks{
	An extended abstract appears in {\em Proc.~13th ACM-SIAM Symp.~on
	Discrete Algorithms,} 2002.}
}

\author{Adam L.~Buchsbaum\thanks{
		AT\&T Labs, Shannon Laboratory,
		180 Park Avenue, 
		Florham Park, NJ 07932, USA, 
		\{alb,gsf\}@research.att.com.}
    \and
	Glenn S.~Fowler$^\dagger$
    \and
	Raffaele Giancarlo\thanks{
		Dipartimento di Matematica ed Applicazioni, 
		Universit\'a di Palermo, Via Archirafi 34, 
		90123 Palermo, Italy, 
		raffaele@altair.math.unipa.it. 
		Work partially supported by AT\&T Labs;
		additional support provided by the
		MURST Project of National Relevance
		Bioinformatica e Ricerca Genomica.}
}

\begin{document}

\maketitle

\begin{abstract}
We study the problem of compressing massive tables
within the partition-training paradigm
introduced by Buchsbaum et al.~[SODA'00],
in which a table is partitioned
by an off-line training procedure
into disjoint intervals of columns,
each of which is compressed separately
by a standard, on-line compressor like gzip.
We provide a new theory
that unifies previous experimental
observations on partitioning
and heuristic observations
on column permutation,
all of which are used to improve compression rates.
Based on the theory,
we devise the first on-line training algorithms
for table compression,
which can be applied to individual files,
not just continuously operating sources;
and also a new, off-line training algorithm,
based on a link to the asymmetric traveling salesman problem,
which improves on prior work 
by rearranging columns prior to partitioning.
We demonstrate these results experimentally.
On various test files,
the on-line algorithms
provide 35--55\% improvement over gzip
with negligible slowdown;
the off-line reordering provides 
up to 20\% further improvement
over partitioning alone.
We also show that a variation
of the table compression problem
is MAX-SNP hard.

\end{abstract}

\section{Introduction}
\label{sec:intro}

\subsection{Table Compression}

Table compression was introduced by Buchsbaum et al.~\cite{pzip:b+00}
as a unique application of compression,
based on several distinguishing characteristics.
Tables are collections of fixed-length records
and can grow to be terabytes in size.
They are often generated by
continuously operating sources
and can contain much redundancy.
An example is a data warehouse at AT\&T that each month
stores one billion records
pertaining to voice phone activity.
Each record is several hundred bytes long
and contains information about
endpoint exchanges,
times and durations of calls,
tariffs,
etc.

The goals of table compression are to be fast, on-line, and effective:
eventual compression ratios of 100:1 or better are desirable.
While storage reduction is an obvious benefit,
perhaps more important is the reduction in subsequent network bandwidth
required for transmission.
Tables of transaction activity,
like phone calls and credit card usage,
are typically stored once but then shipped repeatedly
to different parts of an organization:
for fraud detection,
billing,
operations support, etc.

Prior work \cite{pzip:b+00}
distinguishes tables
from general databases.
Tables are written once and read many times,
while databases are subject to dynamic updates.
Fields in table records are fixed length,
and records tend to be homogeneous;
database records often contain intermixed fixed- and variable-length
fields.
Finally,
the goals of compression differ.
Database compression stresses index preservation,
the ability to retrieve an arbitrary record,
under compression \cite{dbc:c85}.
Tables are typically not indexed at the level of individual
records;
rather, they are scanned in toto by downstream applications.

Consider each record in a table to be a row in a matrix.
A naive method of table compression is to compress the string
derived from scanning the table in row-major order.
Buchsbaum et al.~\cite{pzip:b+00}
observe experimentally that partitioning the table into
contiguous intervals of columns
and compressing each interval separately in this fashion
can achieve significant compression improvement.
The partition is generated by a one-time, off-line training procedure,
and the resulting compression strategy is applied on-line
to the table.
In their application, tables are generated continuously,
so off-line training time can be ignored.
They also observe heuristically that certain rearrangements
of the columns prior to partitioning
further improve compression,
by grouping dependent columns more closely.

We generalize the partitioning approach
into a unified theory
that explains both contiguous partitioning
and column rearrangement.
The theory applies to a set of variables
with a given, abstract notion of combination and cost;
table compression is a concrete case.
To test the theory, we design new algorithms
for contiguous partitioning,
which
speed training to work on-line
on single files in addition to continuously generated tables;
and for reordering in the off-line training
paradigm,
which 
improves the compression rates
achieved from contiguous partitioning alone.
Experimental results support these conclusions.
Before summarizing the results,
we motivate the theoretical insights
by considering the relationship between entropy and compression.

\subsection{Compressive Estimates of Entropy}

Let $\CC$ be a compression
algorithm
and $\CC(x)$
its output on a string $x$.
A large body of work in information
theory establishes the existence of many optimal compression
algorithms:
i.e.,
algorithms such that $|\CC(x)|/|x|$,
the {\em compression rate},
approaches the entropy  of the information source emitting $x$.
These results are usually established via limit theorems, under some
statistical assumptions about the information source. 
For
instance, the LZ77 algorithm
\cite{comp:zl77} is optimal for certain classes of sources,
e.g., stationary and ergodic \cite{it:ct91}.  

While entropy establishes a lower bound on compression rates,
it is not straightforward to measure entropy itself.
One empirical method inverts the relationship
and estimates entropy by applying a provably good compressor
to a sufficiently long, representative string.
That is, the compression rate
becomes a {\em compressive estimate of entropy.}
These estimates themselves become benchmarks
against which future compressors are measured.
Another estimate is the {\em empirical entropy}
of a string,
which is based on the probability distribution
of substrings of various lengths,
without any statistical assumptions regarding the
source emitting the string.
Kosaraju and Manzini \cite{lec:km00}
exploit the synergy between empirical entropy
and true entropy.

The contiguous partitioning approach to table compression
\cite{pzip:b+00}
exemplifies the practical exploitation
of compressive estimates.
Each column of the table
can be seen as being generated by a separate source.
The contiguous partitioning scheme
measures the benefit of a particular partition
empirically,
by compressing the table with respect to that partition
and using the output size as a cost.
Thus,
the partitioning method
uses a compressive estimate
of the joint entropy among columns.
Prior work \cite{pzip:b+00}
demonstrates the benefit of this approach.

\subsection{Method and Results}

We are thus motivated
to study table compression in terms of compressive estimates of the
joint entropy of random variables.  
In Section \ref{sec:InfTh},
we formalize and study two problems on
partitioning sets of variables with abstract
notions of combination and cost;
joint entropy forms one example.
This generalizes the approach of Buchsbaum et al.~\cite{pzip:b+00},
who consider the contiguous case only
and when applied to table compression.
We  develop idealized algorithms
to solve these problems in the general setting.
In Section \ref{sec:optemp},
we apply these methods to table compression
and derive two new algorithms for contiguous partitioning
and one new algorithm for general partitioning
with reordering of columns.
The reordering algorithm
demonstrates a link between general partitioning
and the classical asymmetric traveling salesman problem.
We assess algorithm performance
experimentally in Section \ref{sec:exp}.

The new contiguous partitioning algorithms
are meant to be fast;
better in terms of compression
than off-the-shelf compressors like gzip (LZ77);
but not be as good 
as the optimal, contiguous partitioning algorithm.
The increased training speed 
(compared to optimal, contiguous partitioning) makes the new algorithms
usable in ad hoc settings, however,
when training time must be factored into
the overall time to compress.
We therefore compare compression rates and speeds
to those of gzip and optimal, contiguous partitioning.
For files from various sources,
we achieve 35--55\% improvement in compression
with less than a 1.7-factor slowdown,
both compared to gzip.
For files from genetic databases,
which tend to be harder to compress,
the compression improvement is 5--20\%,
with slowdown factors of 3--8.

The performance of the
general partitioning with reordering algorithm
is predicated on a theorized correlation
between two measures of particular tours in
graphs induced by the compression instances.
We therefore measure this correlation,
and the results
suggest that
the algorithm
is nearly optimal (among partitioning algorithms).
For several of our files,
the 
algorithm yields compression improvements of at least 5\%
compared to optimal, contiguous partitioning
without reordering,
which itself improves over gzip by 20--50\%
for our files.
In some cases, the additional improvement approaches 20\%.
While training time can be ignored in the off-line training paradigm,
we show the additional time for reordering is
not significant.

Finally,
in Sections \ref{sec:hardness}--\ref{sec:rle},
we give some complexity results
that link table compression to the
classical shortest common superstring problem.
We show that an orthogonal (column-major)
variation of table compression is MAX-SNP hard
when LZ77 is the underlying compressor.
On the other hand, while we also show that the row-major
problem is MAX-SNP hard when run length encoding (RLE)
is the underlying compressor,
we prove that the column-major variation
for RLE is solvable in polynomial time.
We conclude with open problems
and directions in Section \ref{sec:conc}.

\section{Partitions of Variables with Entropy-Like Functions}
\label{sec:InfTh}

Let $X=\{x_1,\ldots,x_n\}$ be a set of discrete variables,
each drawn
from some domain $\CD$,
and consider some function $H : \CD^* \rightarrow \Re$.
We use $H(X,Y)$ as a shorthand for $H(Z)$,
where $Z$ is the set composed of all the elements in $X$ and $Y$:
if $X$ and $Y$ are sets, then $Z=X\cup Y$;
if $X$ and $Y$ are variables, then $Z=\{X,Y\}$;
etc.
For some partition $\CP$ of $X$ into subsets,
define $\CH(\CP) = \sum_{Y \in \CP} H(Y)$.
We are interested
in the relationship between $H(X)$ and
$\CH(\CP)$.
For example,
let $X$ be a vector of random variables
with joint probability distribution $p(X)$.
Two vectors $X$ and $Y$
are {\em statistically independent} if and only if $p(x,y)=p(x)p(y)$,
for all $\{x,y\}$;
otherwise, $X$ and $Y$ are {\em statistically dependent}.
Let 
$H(X) = -\sum_{\{x_1,\ldots,x_n\}} p(x_1,\ldots,x_n) \log p(x_1,\ldots,x_n)$
be the {\em joint entropy} of $X$.
Then it is well known \cite{it:ct91}
that for any partition $\CP$ of $X$,
$H(X) \leq \CH(\CP)$,
with equality if and only if all the subsets
in $P$ are mutually independent.

We can also view a table of $n$ columns as a system of $n$
variables.
The relationship between certain compressors
and entropy
suggests that certain rearrangements
that group functionally dependent columns
will lead to better compression;
Buchsbaum et al.~\cite{pzip:b+00}
observe this in practice
while restricting attention to partitions
that preserve the original order of columns.

We are thus motivated to consider generally how
to partition a system of variables
optimally;
i.e., to achieve a partition $\CP$ of $X$
that minimizes $\CH(\CP)$,
for some function $H(\cdot)$,
which we generally call the {\em cost} function.
We introduce the following definitions.
We call an element of $\CP$, which is a subset of $X$, a {\em class}.
We define two variables or sets of variables $X$ and $X'$
to be {\em combinatorially dependent}
if $H(X,X') < H(X) + H(X')$;
otherwise,
$X$ and $X'$ are
{\em combinatorially independent}.
When $H(\cdot)$ is the entropy function over random variables,
combinatorial dependence becomes statistical dependence.
Considering unordered sets implies that
$H(X,X')=H(X',X)$.
Note that in general it is possible that
$H(X,X') > H(X) + H(X')$,
although not when $H(\cdot)$ is the entropy function
over random variables.
Finally,
we define a class
$Y$
to be {\em contiguous}
if $x_i\in Y$ and $x_j \in Y$ for any $i<j$ implies that $x_{i+1}\in Y$
and
a partition $\CP$ to be {\em contiguous} if each $Y\in \CP$ is contiguous.
We now define two problems of finding optimal partitions of $T$.

\begin{problem}
\label{prob:dp}
Find a contiguous
partition $\CP$ of $X$
minimizing $\CH(\CP)$ among all such partitions.
\end{problem}

\begin{problem}
\label{prob:tsp}
Find a 
partition $\CP$ of $X$
minimizing $\CH(\CP)$ among all partitions.
\end{problem}

Clearly,
a solution to Problem \ref{prob:tsp}
is at least as good in terms of cost
as one to Problem \ref{prob:dp}.
Problem \ref{prob:dp}
has a simple, fast algorithmic solution,
however.
Problem \ref{prob:tsp},
while seemingly intractable,
has an
algorithmic heuristic that
seems to work well in practice.

Assume first that combinatorial dependence is
an equivalence relation
on $X$.
This is not necessarily true in practice,
but we study the idealized case
to provide some intuition
for handling real instances,
when we cannot determine combinatorial dependence
or even calculate the true cost function
directly.

\begin{lemma}
\label{lem:clent}
If combinatorial dependence is an equivalence relation
on $X$,
then the partition $\CP$
of $X$ into equivalence classes
$C_1,\ldots,C_k$
solves Problem \ref{prob:tsp}.
\end{lemma}
\begin{proof}
Consider some partition $\CP'\not=\CP$;
we show that $\CH(\CP) \leq \CH(\CP')$.
Assume there exists a class $C'\in \CP'$ such that $C' \supset C_i$
for some $1\leq i\leq k$.  
Partition $C'$ into subclasses $C'_1,\ldots,C'_{\ell}$
such that for each $C'_j$ there is some $C_i$
such that $C'_j \subseteq C_i$.
Let $\CP'' = (\CP' \setminus \{C'\}) \cup \{C'_1,\ldots,C'_{\ell}\}$.
Since the $C_i$'s are equivalence classes,
the $C'_j$'s are mutually independent,
so $H(C') \geq \sum_{j=1}^{\ell} H(C'_j)$,
which implies
$\CH(\CP'') \leq \CH(\CP')$.
Set $\CP' \leftarrow \CP''$,
and
iterate until no such $C'$ exists in $\CP'$.

If no such $C'$ exists in $\CP'$,
then either $\CP' = \CP$, and we are done,
or
else $\CP'$ contains two classes $C'$ and $D'$
such that $C' \cup D' \subseteq C_i$ for some $i$.
The elements in $C'$ and $D'$
are mutually dependent,
so
$H(C', D') < H(C') + H(D')$.
Unite each such pair of classes until $\CP' = \CP$.
\end{proof}

Lemma \ref{lem:clent}
gives a simple algorithm for solving Problem \ref{prob:tsp}
when combinatorial dependence is an equivalence relation
that can be computed:
partition $X$ according to the induced equivalence classes.
When combinatorial dependence is not an equivalence relation,
or when we can only calculate $H(\cdot)$ heuristically,
we seek other approaches.

\subsection{Solutions Without Reordering}
\label{sec:dp}

In the general case,
irrespective of whether combinatorial dependence
is an equivalence relation,
we can solve Problem \ref{prob:dp} 
by dynamic programming.
Let $E[i]$ be the cost of an 
optimal, contiguous partition
of variables $x_1,\ldots,x_i$.
$E[n]$ is thus the cost of a solution
to Problem \ref{prob:dp}.
Define $E[0] = 0$;
then, for $1\leq i \leq n$,

\begin{equation}
\label{eq:dp}
 E[i]= \min_{ 0 \leq j <i} E[j]+H(x_{j+1},\ldots,x_i).
\end{equation}

\noindent
The actual partition with cost $E[n]$ can be maintained
by standard dynamic programming backtracking.

If combinatorial dependence actually is an equivalence relation
and all dependent variables appear contiguously in $X$,
a simple greedy algorithm also solves the problem.
Start with class $C_1 = \{x_1\}$.
In general, let $i$ be the index of the current class
and $j$ be the index of the variable most recently
added to $C_i$.
While $j < n$, iterate as follows.
If $H(C_i \cup \{x_{j+1}\}) < H(C_i) + H(x_{j+1})$,
then set $C_i \leftarrow C_i \cup \{x_{j+1}\}$;
otherwise, start a new class, $C_{i+1} = \{x_{j+1}\}$.
An alternative algorithm
assigns, for $1\leq i < n$,
$x_i$ and $x_{i+1}$ to the same class
if and only if $H(x_i,x_{i+1}) < H(x_i) + H(x_{i+1})$.
We call the resulting partition a greedy partition;
formally, a {\em greedy partition} is one in which
each class is a maximal, contiguous set of mutually dependent variables.

\begin{lemma}\label{le:greedy}
If combinatorial dependence is an equivalence relation and
all combinatorially dependent variables appear contiguously in $X$,
then the greedy partition solves
Problems \ref{prob:dp} and \ref{prob:tsp}.
\end{lemma}
\begin{proof}
By assumption,
the classes in a greedy partition
correspond to the equivalence classes of $X$.
Lemma \ref{lem:clent}
thus shows that the greedy partition
solves Problem \ref{prob:tsp}.
Contiguity therefore implies
it also solves Problem \ref{prob:dp}.
\end{proof}

\subsection{Solutions with Reordering}
\label{sec:tsp}

Problem  \ref{prob:tsp}
asks for the best way to partition the variables in $T$,
ignoring contiguity constraints.
While a general solution seems intractable,
we give a combinatorial approach that admits
a practical heuristic.

Define a weighted, complete, undirected graph, $G(X)$,
with a vertex for each $x_i\in X$;
the {\em weight} of edge $\{x_i,x_j\}$
is $w(x_i,x_j) = \min(H(x_i,x_j), H(x_i) + H(x_j))$.
Let $P=(v_0,\ldots,v_{\ell})$ be any path in $G(X)$.
The {\em weight} of $P$ is $w(P) = \sum_{i=0}^{\ell-1}w(v_i,v_{i+1})$.
We apply the cost function $H(\cdot)$
to define the cost of $P$.
Consider removing all edges $\{u,v\}$ from $P$
such that $u$ and $v$ are combinatorially independent.
This leaves a set of disjoint paths,
$\CS(P) = \{P_1,\ldots,P_k\}$ for some $k$.
We define the {\em cost} of $P$ to
be $\CH(P) = \sum_{i=1}^k H(P_i)$,
where $P_i$ is taken to be the unordered set
of vertices in the corresponding subpath.
If $P$ is a tour of $G(X)$,
then $\CS(P)$ corresponds to a partition of $X$.

We establish a relationship between the cost and weight
of a tour $P$.
Assume there are two distinct paths
$P_i = (u_0,\ldots,u_k)$
and
$P_j = (v_0,\ldots,v_{\ell})$
in $\CS(P)$
such that 
$u_k$ and $v_0$
are combinatorially dependent
and $v_0$ follows $u_k$ in $P$.
In $P$ exist the edges $\{u_k,x\}$,
$\{y,v_0\}$,
and $\{v_{\ell},z\}$.
We can transform $P$ into a new tour $P'$
that
unites $P_i$ and $P_j$
by substituting for these three edges
the new edges:
$\{u_k,v_0\}$,
$\{v_{\ell},x\}$,
and $\{y,z\}$.
We call this a {\em path coalescing transformation}.
The following lemma shows that it is a restricted form
of the standard traveling salesman 3-opt transformation,
in that it always reduces the cost of a tour.
It is restricted by the stipulation that $u_k$ and $v_0$
be combinatorially dependent.

\begin{lemma}
\label{lem:costless}
If $P'$ is formed from $P$ by a path coalescing transformation,
then
$w(P') < w(P)$.
\end{lemma}
\begin{proof}
Consider
\begin{equation}
\label{eq:old}
	w(u_k,x) + w(y,v_0) + w(v_{\ell},z)
\end{equation}
and
\begin{equation}
\label{eq:new}
	w(u_k,v_0) + w(v_{\ell},x) + w(y,z).
\end{equation}
We have $w(P') - w(P) = \equ{eq:new} - \equ{eq:old}$.
The definition of $\CS(P)$
implies that 
$\equ{eq:old} = H(u_k) + H(x) + H(y) + H(v_0) + H(v_{\ell}) + H(z)$.
That $u_k$ and $v_0$ are combinatorially dependent
implies $w(u_k,v_0) <  H(u_k)+ H(v_0)$.
Since $w(X,Y) \leq H(X) + H(Y)$ for any $X$ and $Y$,
we conclude that $\equ{eq:new} < \equ{eq:old}$.
\end{proof}

Repeated path coalescing groups combinatorially dependent
variables.  
If a tour $P$ admits no path coalescing transformation,
and if combinatorial dependence is an equivalence relation on $X$,
then we can conclude that $P$ is optimal
by Lemma \ref{lem:clent}.
That is, $\CS(P)$ corresponds to an optimal partition
of $X$, which solves Problem \ref{prob:tsp}.
Furthermore,
Lemma \ref{lem:costless}
implies that a minimum weight tour
$P$ admits no path coalescing transformation.

When
$H(\cdot)$ is sub-additive, i.e., $H(X,Y) \leq H(X)+H(Y)$, as is the
entropy function, a sequence of path coalescing transformations
yields a sequence of paths of non-increasing costs. That is,
in Lemma \ref{lem:costless}, $w(P')<w(P)$ {\em and} $\CH(P') \leq \CH(P)$.  We
explore this connection between the two
functions below, when we do not assume that combinatorial dependence is an
equivalence relation or even that $H(\cdot)$ is sub-additive.

\section{Partitions of Tables and Compression}
\label{sec:optemp}

We apply the results of Section \ref{sec:InfTh}
to table compression.
Let $T$ be a table of $n=|T|$ columns
and some fixed, arbitrary number of rows.
Let $T[i]$ denote the $i$'th column of $T$.
Given two tables $T_1$ and $T_2$,
let $T_1T_2$ be the table formed by their juxtaposition.
That is, $T=T_1T_2$ is defined so that
$T[i] = T_1[i]$ for $1\leq i\leq |T_1|$
and $T[i] = T_2[i-|T_1|]$ for $|T_1| < i \leq |T_1| + |T_2|$.
Any column is a one-column table,
so $T[i]T[j]$ is the table formed by projecting
the $i$'th and $j$'th columns of $T$;
and so on.
We use the shorthand $T[i,j]$
to represent the projection $T[i]\cdots T[j]$
for some $j\geq i$.

Fix a compressor $\CC$:
e.g.,
gzip, based on LZ77 \cite{comp:zl77};
compress, based on LZ78 \cite{comp:w84,comp:zl78};
or bzip, based on Burrows-Wheeler \cite{comp:bw94}.
Let $\HC(T)$ be the size of the result of compressing
table $T$ as a string in row-major order
using \CC.
Let $\HC(T_1,T_2) = \HC(T_1T_2)$.
$\HC(\cdot)$
is a cost function as discussed in Section \ref{sec:InfTh},
and the definitions of combinatorial dependence
and independence apply to tables.
In particular,
two tables $T_1$ and $T_2$,
which might be projections of columns from a common table $T$,
are {\em combinatorially dependent}
if $\HC(T_1,T_2) < \HC(T_1) + \HC(T_2)$---if
compressing them together is better
than compressing them separately---and
{\em combinatorially independent} otherwise.

Problems \ref{prob:dp} and \ref{prob:tsp}
now apply to compressing $T$.
Problem \ref{prob:dp}
is to find a contiguous partition of $T$
into intervals of columns
minimizing
the overall cost of compressing each interval separately.
Problem \ref{prob:tsp}
is to find a partition of $T$,
allowing columns to be reordered,
minimizing
the overall cost of compressing each interval separately.
Buchsbaum et al.~\cite{pzip:b+00}
address Problem \ref{prob:dp} experimentally
and leave Problem \ref{prob:tsp} open save
for some heuristic observations.

A few major issues arise in this application.
Combinatorial dependence is not necessarily an equivalence relation.
It is not necessarily even symmetric,
so we can no longer ignore the order
of columns in a class.
Also,
$\HC(\cdot)$ need not be sub-additive.
If \CC~behaves according to entropy, however,
then
intuition suggests that our partitioning strategies
will improve compression.
Stated conversely,
if $\HC(T)$ is far from $H(T)$,
the entropy of $T$,
there should be some partition $P$
of
$T$ so that $\HC(P)$ approaches $H(T)$,
which is a lower bound on $\HC(T)$.
We will present algorithms
for solving these problems
and experiments assessing
their performance.

\subsection{Algorithms for Table Compression without Rearrangement of Columns}
\label{sec:tablecalg}

The dynamic programming solution
in Equation \equ{eq:dp}
finds an optimal, contiguous partition
solving Problem \ref{prob:dp}.
Buchsbaum et al.~\cite{pzip:b+00}
demonstrate experimentally
that it effectively improves compression results,
and we will use their method as a benchmark.

The dynamic program,
however, requires $\Theta(n^2)$
steps,
each applying \CC~to an average
of $\Theta(n)$ columns,
for a total of $\Theta(n^3)$ column compressions.
In the off-line training paradigm,
this optimization time can be ignored.
Faster algorithms,
however,
might allow some partitioning to be applied
when compressing single, tabular files
in addition to continuously generated tables.

The greedy algorithms from Section \ref{sec:dp}
apply directly in our framework.
We denote by GREEDY the algorithm 
that grows class $C_i$ incrementally
by comparing $\HC(C_iT[j+1])$ and
$\HC(C_i) + \HC(T[j+1])$.
We denote by GREEDYT the algorithm
that assigns $T[i]$ and $T[i+1]$
to the same class
when $\HC(T[i,i+1]) < \HC(T_i) + \HC(T[i+1])$.

GREEDY performs $2(n-1)$ compressions, each of
$\Theta(n)$ columns,
for a total of $\Theta(n^2)$ column compressions.
GREEDYT performs $2(n-1)$ compressions,
each of one or two columns,
for a total of $\Theta(n)$ column compressions,
asymptotically at least as fast as applying $\CC$ to $T$ itself.

Even though combinatorial dependence is not an equivalence relation,
we hypothesize that GREEDY and GREEDYT will 
produce partitions close in cost to the optimal contiguous
partition produced by the dynamic program.
We present experimental results
testing this hypothesis in Section \ref{sec:exp}.

\subsection{Algorithms for Table Compression with  Rearrangement of Columns}

We now consider Problem \ref{prob:tsp}.
Assuming that combinatorial dependence is not an equivalence relation,
to the best of our
knowledge, the only known algorithm to solve it exactly consists of
generating all $n!$ column orderings and applying the dynamic program
in Equation \equ{eq:dp} to each.
The relationship between compression and entropy,
however, suggests that the approach in Section \ref{sec:tsp}
can still be fruitfully applied.

Recall that in the idealized case,
an optimal solution corresponds to a tour
of $G(T)$
that admits no path coalescing transformation.
Furthermore,
such transformations
always reduce the weight of such tours.
The lack of symmetry in $\HC(\cdot)$
further suggests that order within classes
is important:
it no longer suffices to coalesce paths globally.

We therefore hypothesize a strong, positive correlation
between tour
weight and compression cost.
This would imply that a traveling salesman (TSP) tour of $G(T)$
would yield an optimal or near-optimal partition of $T$.
To test this hypothesis,
we generate a set of tours of various weights,
by iteratively applying standard optimizations (e.g., 3-opt, 4-opt).
Each tour induces an ordering of the columns,
which
we optimally partition using the dynamic program.
We present results of this experiment in Section \ref{sec:exp}.

\section{Experiments}
\label{sec:exp}

\subsection{Data}

We report experimental results on several data sets.
The first three of the following
are used by Buchsbaum et al.~\cite{pzip:b+00}.

\begin{description}
\item[{\sc care}] is a collection of 90-byte records
from a customer care database
of voice call activity.
\item[{\sc network}] is a collection of 32-byte records
from a system of network status monitors.
\item[{\sc census}]
is a portion of the United States
{\em 1990 Census of Population and Housing Summary Tape File 3A}
\cite{uscd92}.
We used field group 301, level 090, for all states.
Each record is 932 byes.
\item[{\sc lerg}] is a file from Telcordia's
database describing local telephone switches.
We appended spaces as necessary
to pad each record to a uniform 30 bytes.
\end{description}

We also use several files from genetic databases,
which are growing at a fast pace
and
pose unique challenges to compression systems
\cite{gsc:gt94,protcomp:nw99}.
These files
can be viewed as two-dimensional, alphanumeric tables
representing
multiple alignments of proteins (amino acid sequences) and
genomic coding regions (DNA sequences).

The files
{\sc EGF}, {\sc LRR}, {\sc PF00032}, {\sc backPQQ}, {\sc callagen}, and {\sc cbs}
come from the
Pfam database of multiple alignments of protein domains or
conserved protein functions \cite{pfam:b+00}.
Its main function is to store
information that can be used to determine whether a new
protein belongs to an existing domain or family. It contains more than
1800 protein families and has many
mirror sites. 
The size of each table can range from a few
columns and rows to hundreds of columns and thousands of rows.
We have chosen multiple alignments of different sizes and
representing protein domains with differing degrees of
conservation:
i.e., how close two members of a family are in terms of 
matching characters in the alignment. 

The file
{\sc cytoB} is from the
AMmtDB database of multi-aligned sequences of
Vertebrate mitochondrial genes for coding proteins \cite{ammtdb:l+00}.
It contains
data from 888 different species and over 1100 multi-alignments of
protein-coding genes. The tables corresponding to the alignments tend
to have rows in the order of hundreds and columns in the order of
thousands,
much wider than the other files we consider. 
We have experimented with one multiple alignment: 
{\sc CytoB} represents the coding region 
of the mitochondrial gene (from 500 different species) of cytocrome B.

Table \ref{tab:desc} details the sizes of the files
and how well gzip and the optimal partition via
dynamic programming
(using gzip as the underlying compressor)
compress them.
We use the pin/pzip system described by Buchsbaum et al.~\cite{pzip:b+00}
to general optimal, contiguous partitions.
For each file,
we run the dynamic program on a small {\em training set}
and compress the remainder of the data, the {\em test set}.
Gzip results are with respect to the test sets
only.
Buchsbaum et al.~\cite{pzip:b+00}
investigate the relationship between training size
and compression performance
and demonstrate a threshold after which
more training data does not improve performance.
Here we simply use enough training data to
exceed this threshold
and report this amount in Table \ref{tab:desc}.
The training and test sets remain disjoint to
support the validity of using a partition
from a small amount of training data
on a larger amount of subsequent data.
In a real application, the training data would also be compressed.

All experiments were performed 
on one 250 MHz R10000 processor
in a 24-processor SGI Challenge,
with 14 GB of main memory.
Each time reported is the medians of five runs.

\begin{table}
\caption{Files used in our experiments.
Bpr is bytes per record.
Size is the original size of the file in bytes.
Training size is the ratio of the size of the training
set to that of the test set.
Gzip and DP report compression results;
DP is the optimal contiguous partition,
calculated by dynamic programming.
For each, Size is the size of the compressed file in bytes,
and Rate is the ratio of compressed to original size.
DP/Gzip shows the relative improvement
yielded by partitioning.}
\label{tab:desc}
\begin{center}
\begin{tabular}{l|rrr|rr|rr|r}
 & & & \multicolumn{1}{c|}{Training} & \multicolumn{2}{c|}{Gzip} & \multicolumn{2}{c|}{DP} & \\
File & \multicolumn{1}{c}{Bpr} & \multicolumn{1}{c}{Size} & \multicolumn{1}{c|}{Size} & \multicolumn{1}{c}{Size} & \multicolumn{1}{c|}{Rate} & \multicolumn{1}{c}{Size} & \multicolumn{1}{c|}{Rate} & \multicolumn{1}{c}{DP/Gzip} \\ \hline
{\sc care} & 90 & 8181810 & 0.0196 & 2036277 & 0.2489 & 1290936 & 0.1578 & 0.6340 \\
{\sc network} & 126  & 60889500 & 0.0207 & 3749625 & 0.0616 & 1777790 & 0.0292 & 0.4741 \\
{\sc census} & 932 & 332959796 & 0.0280 & 30692815 & 0.0922 & 21516047 & 0.0646 & 0.7010 \\
{\sc lerg} & 30  & 3480030 & 0.0862 & 454975 & 0.1307 & 185856 & 0.0534 & 0.4085 \\
{\sc EGF} & 188  & 533920 & 0.0690 & 72305 & 0.1354 & 56571 & 0.1060 & 0.7824 \\
{\sc LRR} & 72  & 235440 & 0.0685 & 61745 & 0.2623 & 49053 & 0.2083 & 0.7944 \\
{\sc PF00032} & 176  & 402512 & 0.0673 & 34225 & 0.0850 & 30587 & 0.0760 & 0.8937 \\
{\sc backPQQ} & 81  & 22356 & 0.0507 & 7508 & 0.3358 & 7186 & 0.3214 & 0.9571 \\
{\sc callagen} & 112  & 242816 & 0.0678 & 67338 & 0.2773 & 59345 & 0.2444 & 0.8813 \\
{\sc cbs} & 134  & 73834 & 0.0635 & 23207 & 0.3143 & 19839 & 0.2687 & 0.8549 \\
{\sc cytoB} & 1225  & 579425 & 0.0592 & 109681 & 0.1893 & 89983 & 0.1553 & 0.8204
\end{tabular}
\end{center}
\end{table}

\subsection{Greedy Algorithms}

Our hypothesis 
that GREEDY and GREEDYT produce partitions
close in cost to that of the optimal, contiguous partition,
if true implies that
we can substitute the greedy algorithms for the dynamic
program (DP) in purely on-line applications
that cannot afford off-line training time.
We thus compare compression rates
of GREEDY and GREEDYT against DP and gzip,
to assess the quality of the partitions;
and we compare the time taken by GREEDY and GREEDYT
(partitioning and compression)
against gzip,
to assess tractability.
Table \ref{tab:greedysize} shows the resulting compressed sizes
using partitions computed with GREEDY and GREEDYT.
Table \ref{tab:greedytime} gives the time results.

\begin{table}
\caption{Performance of GREEDY and GREEDYT.
For each, Size is the size of the compressed file
using the corresponding partition;
Rate is the corresponding compression rate;
/Gzip is the size relative to gzip;
and /DP is the size relative to
using the optimal, contiguous partition.}
\label{tab:greedysize}
\begin{center}
\begin{tabular}{l|rrrr|rrrr}
 & \multicolumn{4}{c|}{GREEDY} & \multicolumn{4}{c}{GREEDYT} \\
File & 
\multicolumn{1}{c}{Size} &
\multicolumn{1}{c}{Rate} &
\multicolumn{1}{c}{/Gzip} & 
\multicolumn{1}{c|}{/DP} &
\multicolumn{1}{c}{Size} &
\multicolumn{1}{c}{Rate} &
\multicolumn{1}{c}{/Gzip} &
\multicolumn{1}{c}{/DP} \\ \hline
{\sc care} & 1307781 & 0.1598 & 0.6422 & 1.0130 & 1360160 & 0.1662 & 0.6680 & 1.0536 \\
{\sc network} & 1784625 & 0.0293 & 0.4759 & 1.0038 & 2736366 & 0.0449 & 0.7298 & 1.5392 \\
{\sc census} & 21541616 & 0.0647 & 0.7018 & 1.0012 & 21626399 & 0.0650 & 0.7046 & 1.0051 \\
{\sc lerg} & 197821 & 0.0568 & 0.4348 & 1.0644 & 199246 & 0.0573 & 0.4379 & 1.0720 \\
{\sc EGF} & 57016 & 0.1068 & 0.7885 & 1.0079 & 61178 & 0.1146 & 0.8461 & 1.0814 \\
{\sc LRR} & 49778 & 0.2114 & 0.8062 & 1.0148 & 49393 & 0.2098 & 0.8000 & 1.0069 \\
{\sc PF00032} & 31037 & 0.0771 & 0.9069 & 1.0147 & 31390 & 0.0780 & 0.9172 & 1.0263 \\
{\sc backPQQ} & 7761 & 0.3472 & 1.0337 & 1.0800 & 7761 & 0.3472 & 1.0337 & 1.0800 \\
{\sc callagen} & 58952 & 0.2428 & 0.8755 & 0.9934 & 56313 & 0.2319 & 0.8363 & 0.9489 \\
{\sc cbs} & 21571 & 0.2922 & 0.9295 & 1.0873 & 21939 & 0.2971 & 0.9454 & 1.1059 \\
{\sc cytoB} & 94128 & 0.1625 & 0.8582 & 1.0461 & 113160 & 0.1953 & 1.0317 & 1.2576
\end{tabular}
\end{center}
\end{table}

\begin{table}
\begin{center}
\caption{On-line performance of GREEDY and GREEDYT.
For each, Time is the time in seconds
to compute the partition and compress the file;
/Gzip is the time relative to gzip.}
\label{tab:greedytime}
\begin{tabular}{l|r|rr|rr}
 & \multicolumn{1}{c|}{Gzip} & \multicolumn{2}{c|}{GREEDY} & \multicolumn{2}{c}{GREEDYT} \\
File & 
\multicolumn{1}{c|}{Time} &
\multicolumn{1}{c}{Time} &
\multicolumn{1}{c|}{/Gzip} &
\multicolumn{1}{c}{Time} &
\multicolumn{1}{c}{/Gzip} \\ \hline
{\sc care} & 5.0260 & 7.1020 & 1.4131 & 6.4340 & 1.2801 \\
{\sc network} & 15.0000 & 25.3790 & 1.6919 & 24.2750 & 1.6183 \\
{\sc census} & 126.6450 & 160.7960 & 1.2697 & 147.1980 & 1.1623 \\
{\sc lerg} & 1.5730 & 2.2800 & 1.4495 & 2.3080 & 1.4673 \\
{\sc EGF} & 0.2350 & 0.8030 & 3.4170 & 0.7250 & 3.0851 \\
{\sc LRR} & 0.1260 & 0.4530 & 3.5952 & 0.4450 & 3.5317 \\
{\sc PF00032} & 0.1320 & 0.8950 & 6.7803 & 0.6290 & 4.7652 \\
{\sc backPQQ} & 0.0180 & 0.3090 & 17.1667 & 0.3260 & 18.1111 \\
{\sc callagen} & 0.2500 & 0.6050 & 2.4200 & 0.5300 & 2.1200 \\
{\sc cbs} & 0.0530 & 0.4260 & 8.0377 & 0.4020 & 7.5849 \\
{\sc cytoB} & 0.8230 & 3.7330 & 4.5358 & 2.1830 & 2.6525
\end{tabular}
\end{center}
\end{table}

GREEDY compresses to within 2\% of DP
on seven of the files,
including four of the genetic files.
It is never more than 9\% bigger than DP,
and with the exception of {\sc backPQQ},
always outperforms gzip.
GREEDYT comes within 10\% of DP on seven files,
including four genetic files
and outperforms gzip except on {\sc backPQQ} and {\sc cytoB}.
Both GREEDY and GREEDYT seem to outperform DP on {\sc callagen},
although this would seem theoretically impossible.
It is an artifact of the training/testing paradigm:
we compress data distinct
from that used to build the partitions.

Tables \ref{tab:greedysize} and \ref{tab:greedytime} show that in many cases,
the greedy algorithms provide significant extra compression
at acceptable time penalties.
For the non-genetic files,
greedy partitioning compression
is less than $1.7$ slower than gzip
yet provides 35--55\% more compression.
For the genetic files,
the slowdown is a factor of 3--8,
and the extra compression is 5--20\%
(ignoring {\sc backPQQ}).
Thus, the greedy algorithms
provide a good on-line heuristic
for improving compression.

\subsection{Reordering via TSP}

Our hypothesis that tour weight and compression are correlated
implies that generating a TSP tour (or approximation)
would yield an optimal (or near optimal) partition.
Although we do not know what the optimal partition
is for our files,
we can assess the correlation
by generating a sequence of tours and,
for each,
measuring the resulting compression.
We also compare the compression
using the best partition
from the sequence against that
using DP on the original ordering,
to gauge the improvement yielded by reordering.

\begin{figure}
\figinbox{3in}{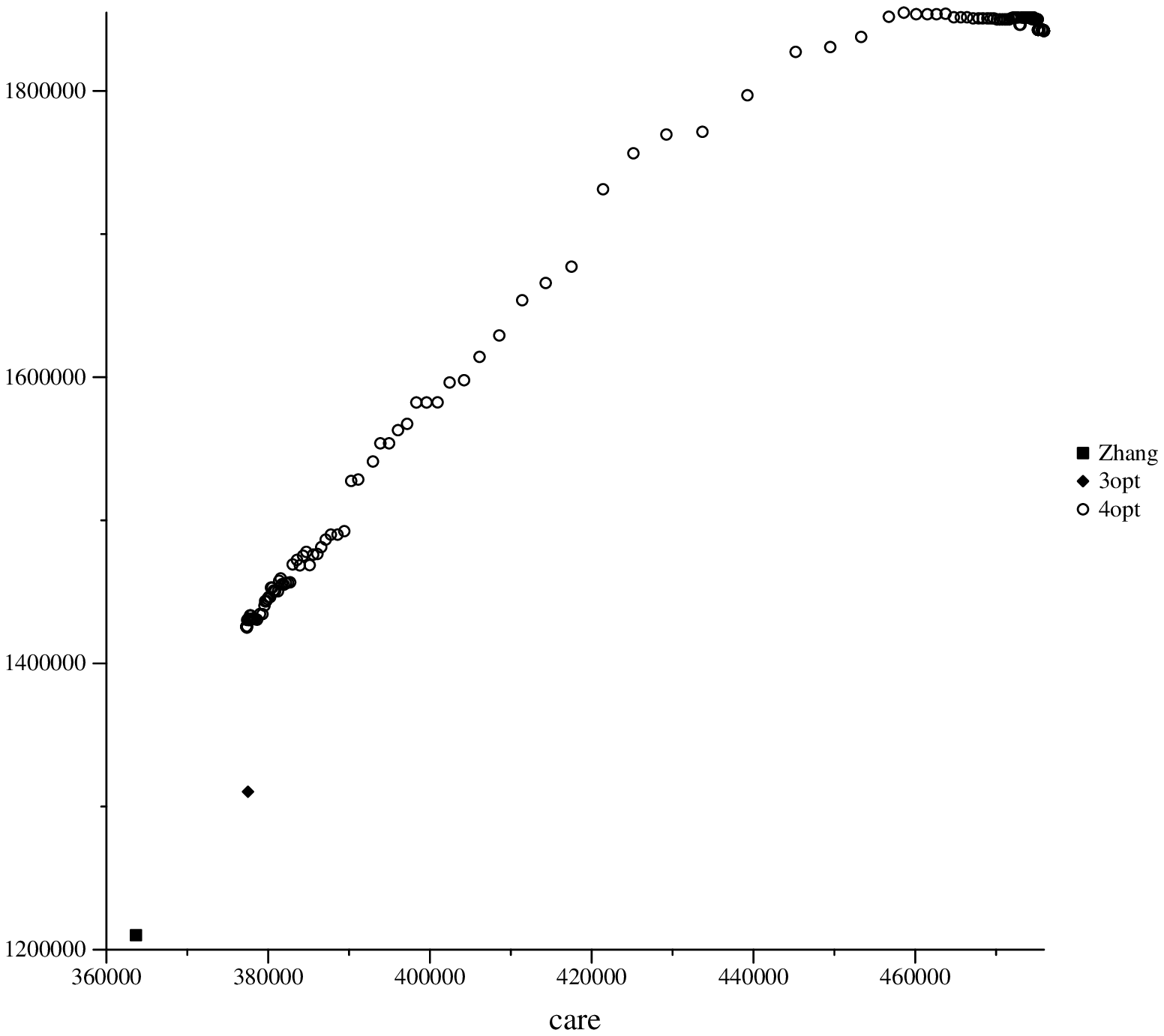}
\hfill
\figinbox{3in}{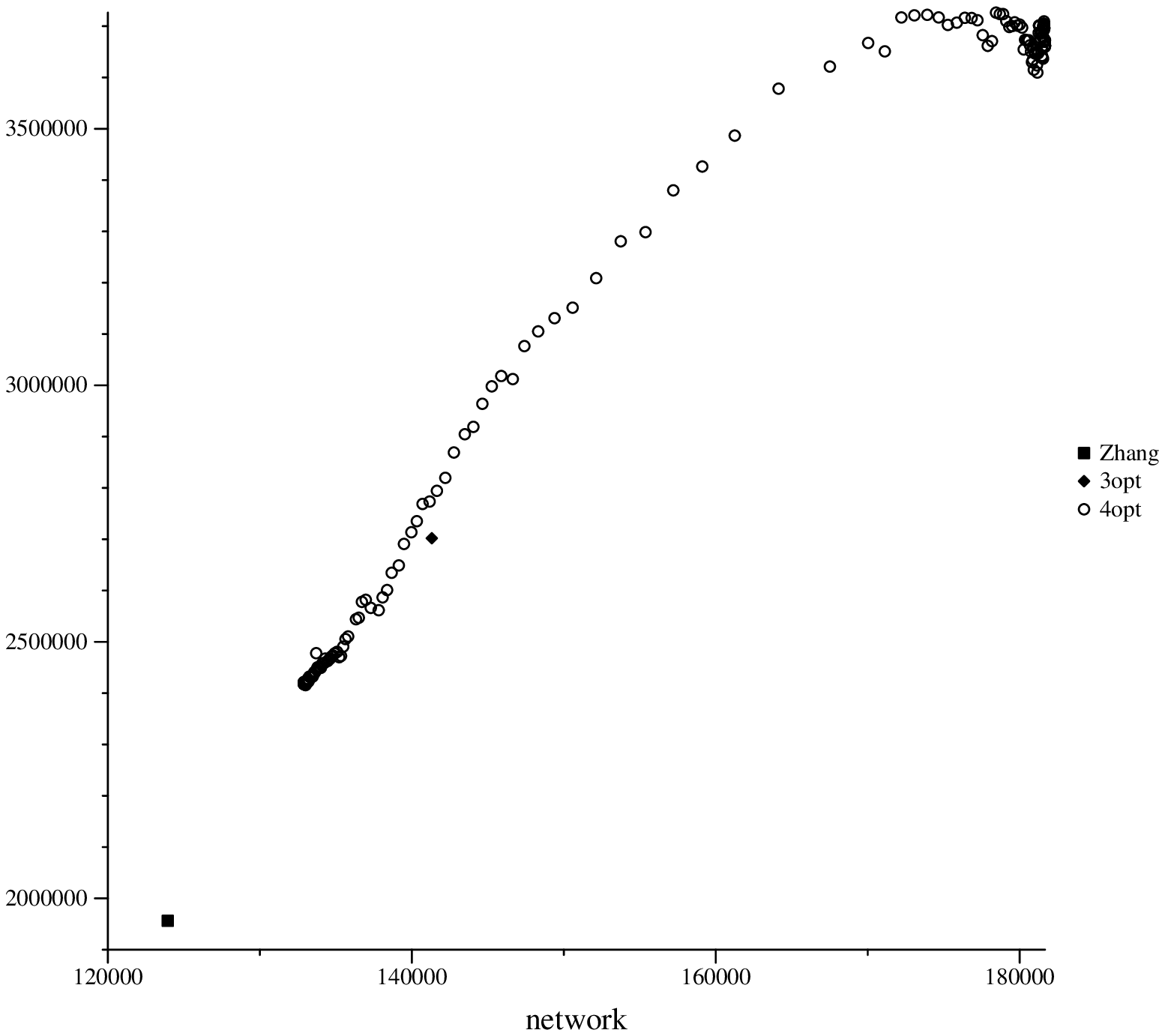}
\figinbox{3in}{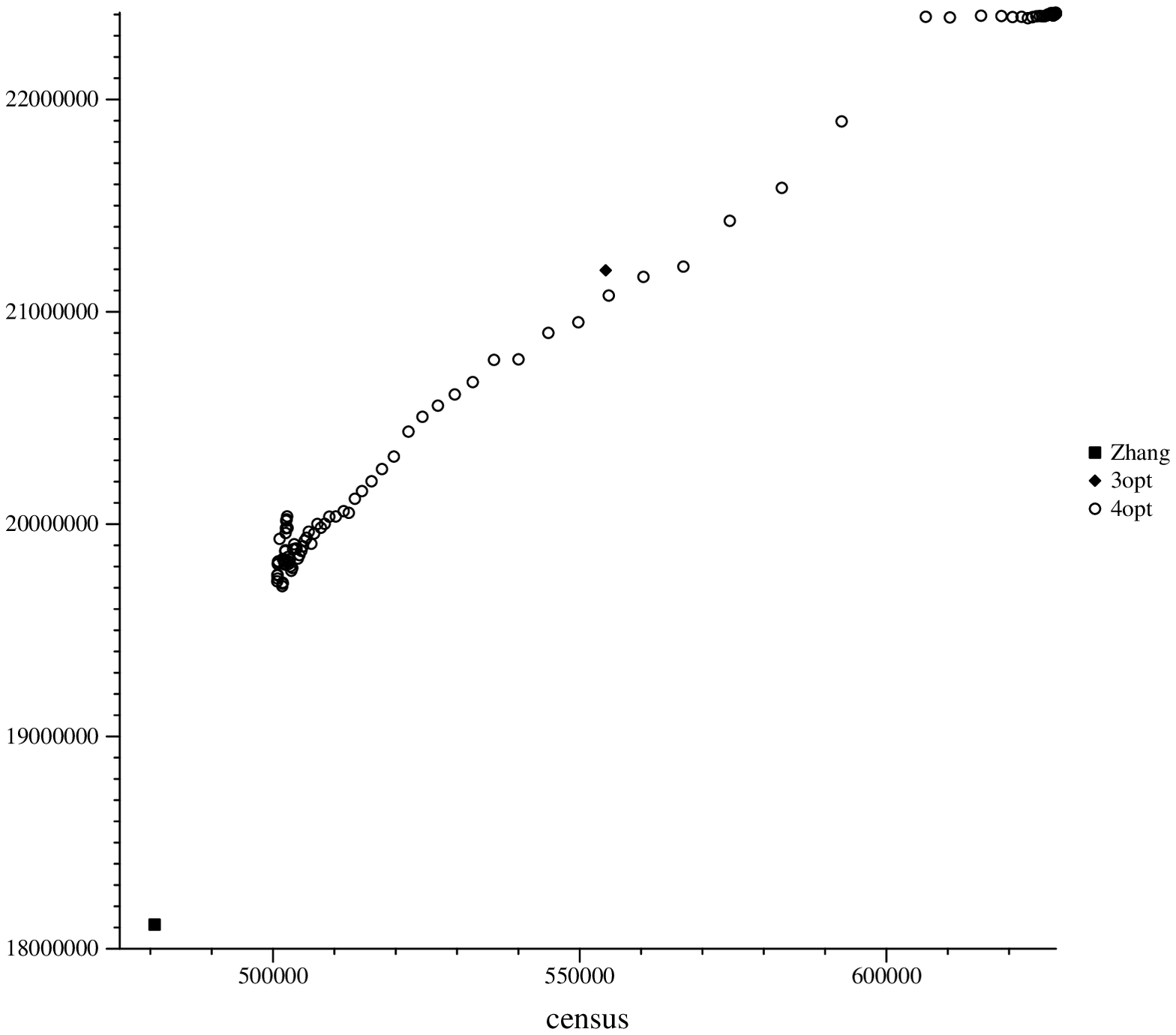}
\hfill
\figinbox{3in}{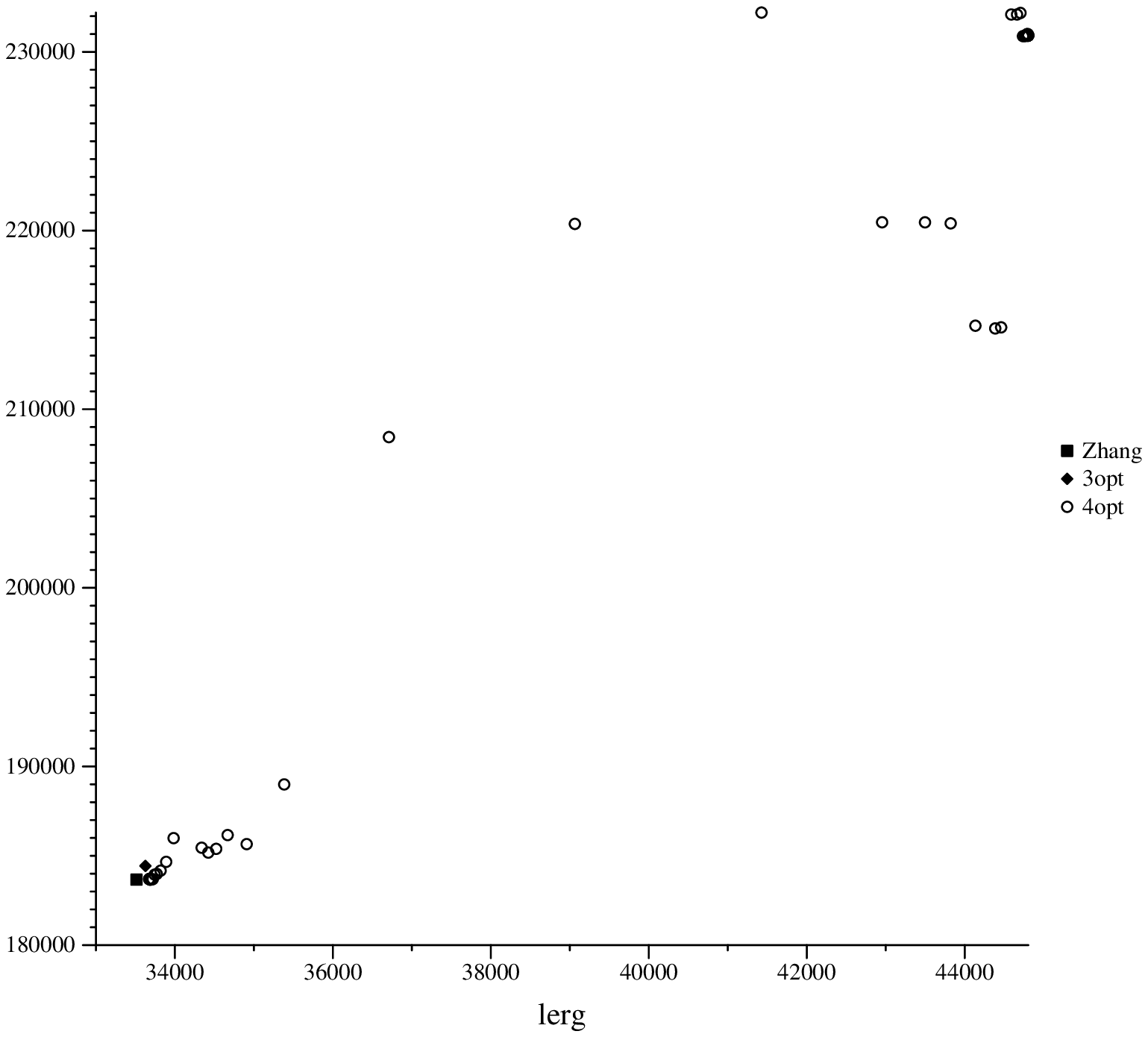}
\figinbox{3in}{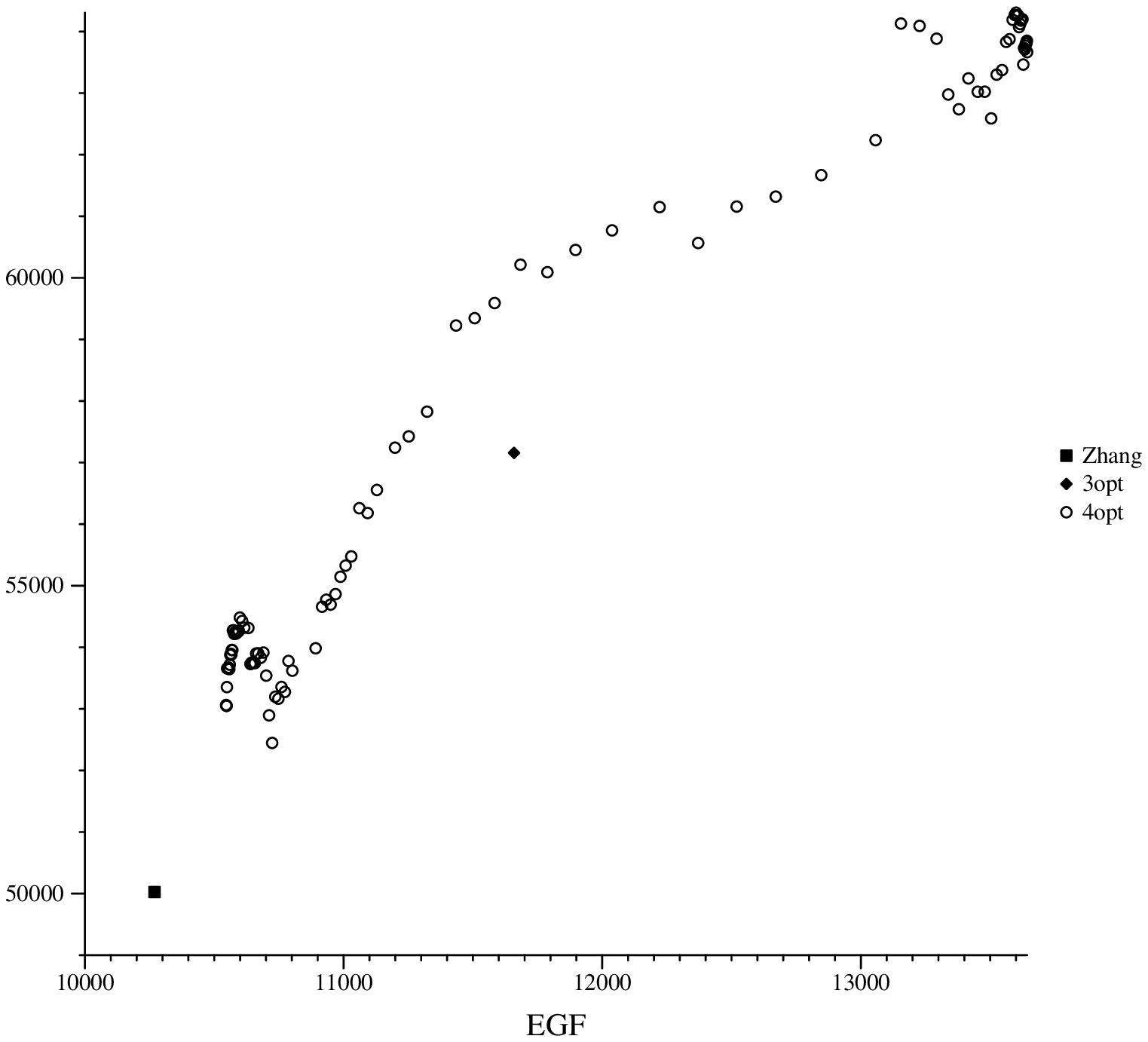}
\hfill
\figinbox{3in}{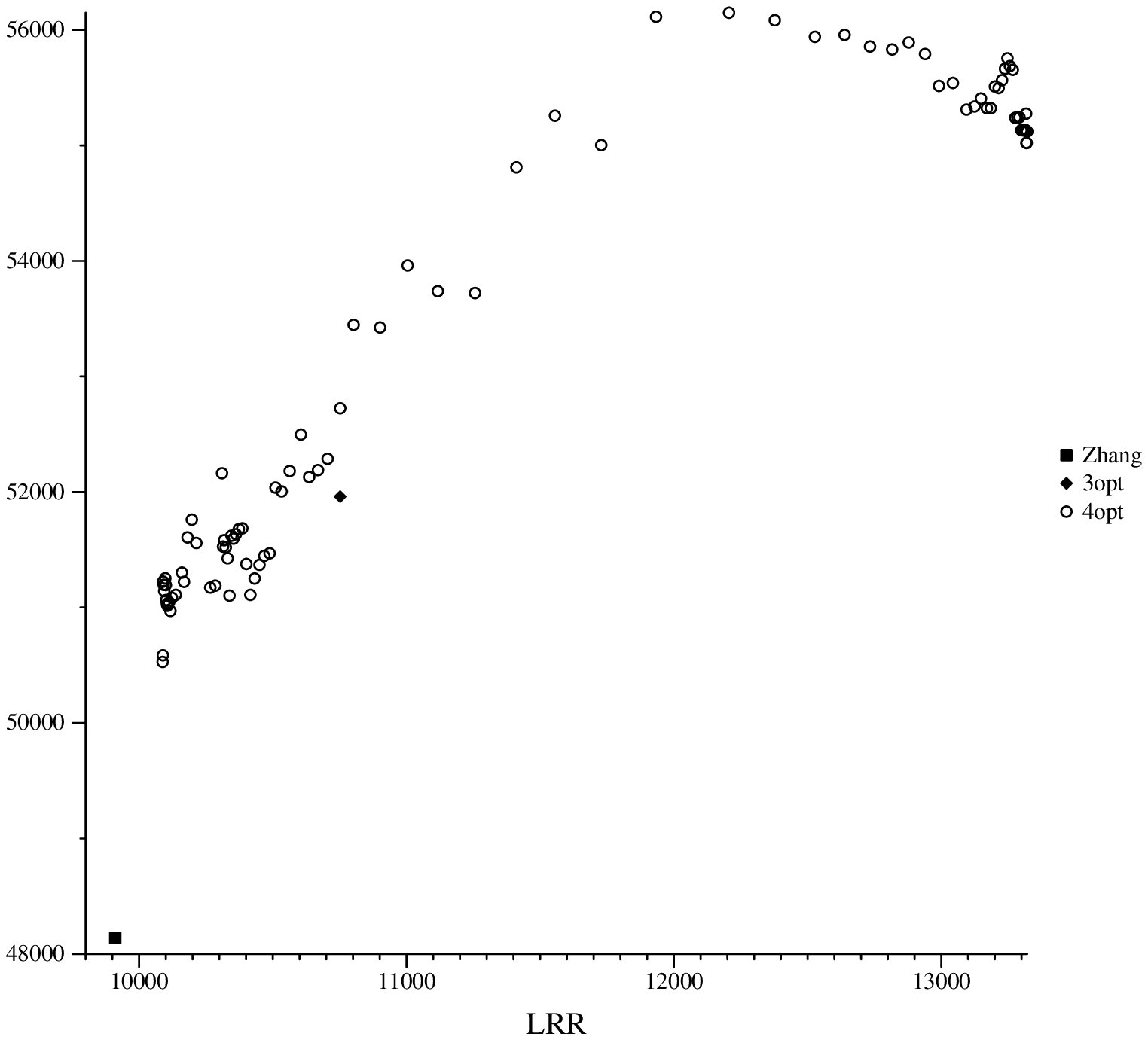}
\caption{Relationship between tour cost (x-axes) and compression
size (y-axes) for {\sc care}, {\sc network}, {\sc census}, {\sc lerg}, {\sc EGF}, and {\sc LRR},
using the result of Zhang's algorithm,
a 3-opt local optimum,
and a sequence of tours
generated by a series of 4-opt changes.}
\label{fig:tsp1}
\end{figure}

\begin{figure}
\figinbox{3in}{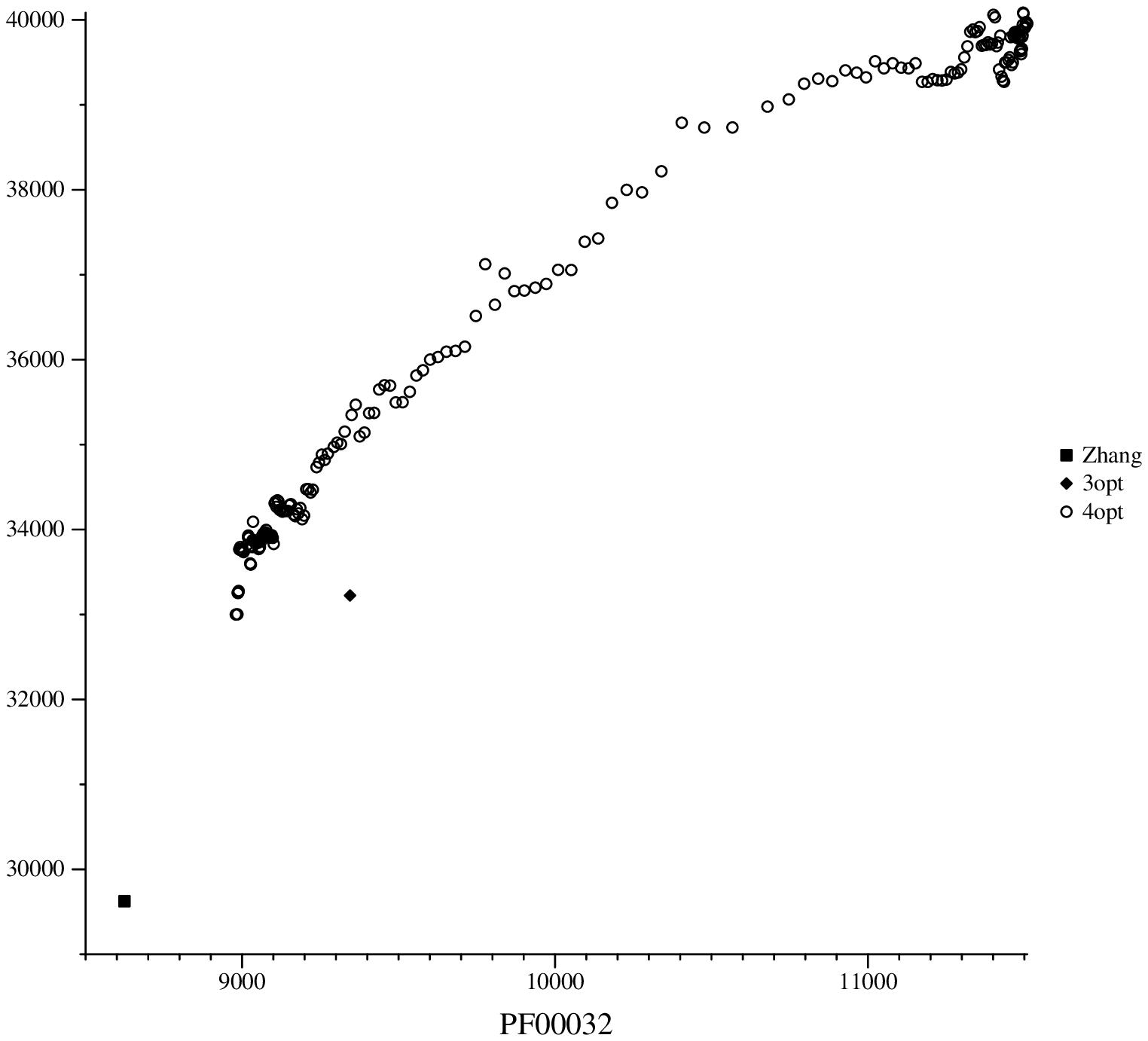}
\hfill
\figinbox{3in}{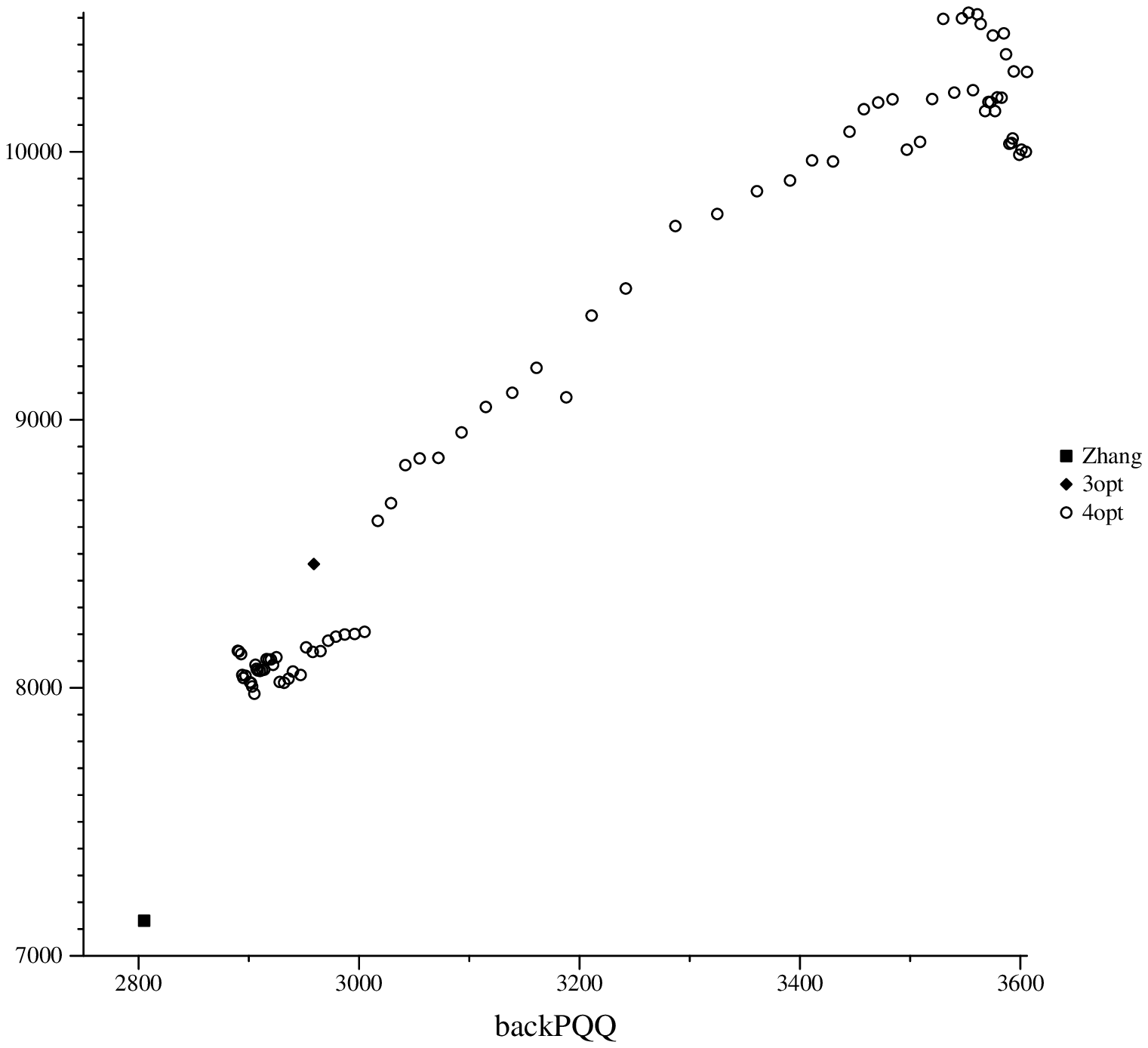}
\figinbox{3in}{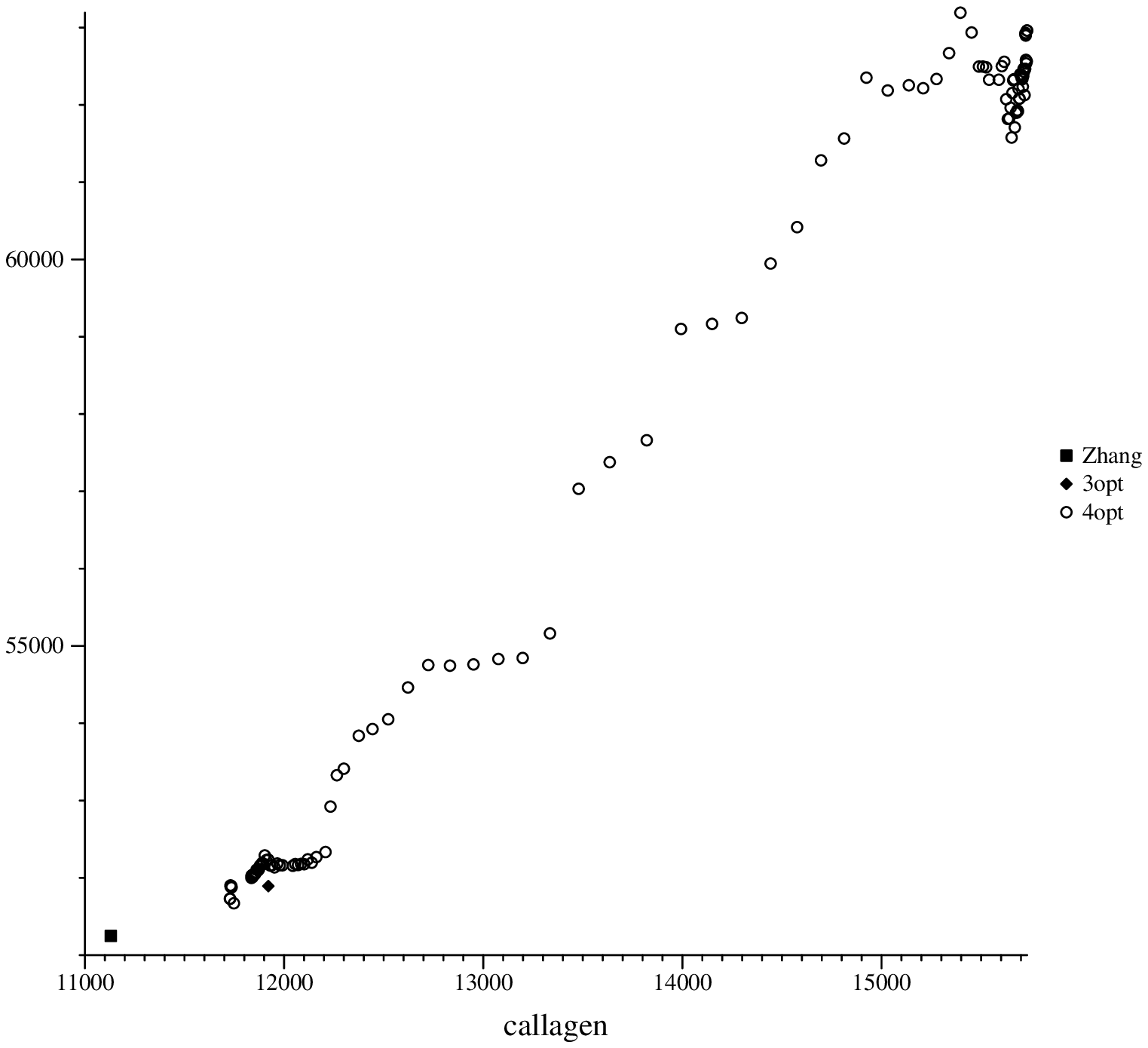}
\hfill
\figinbox{3in}{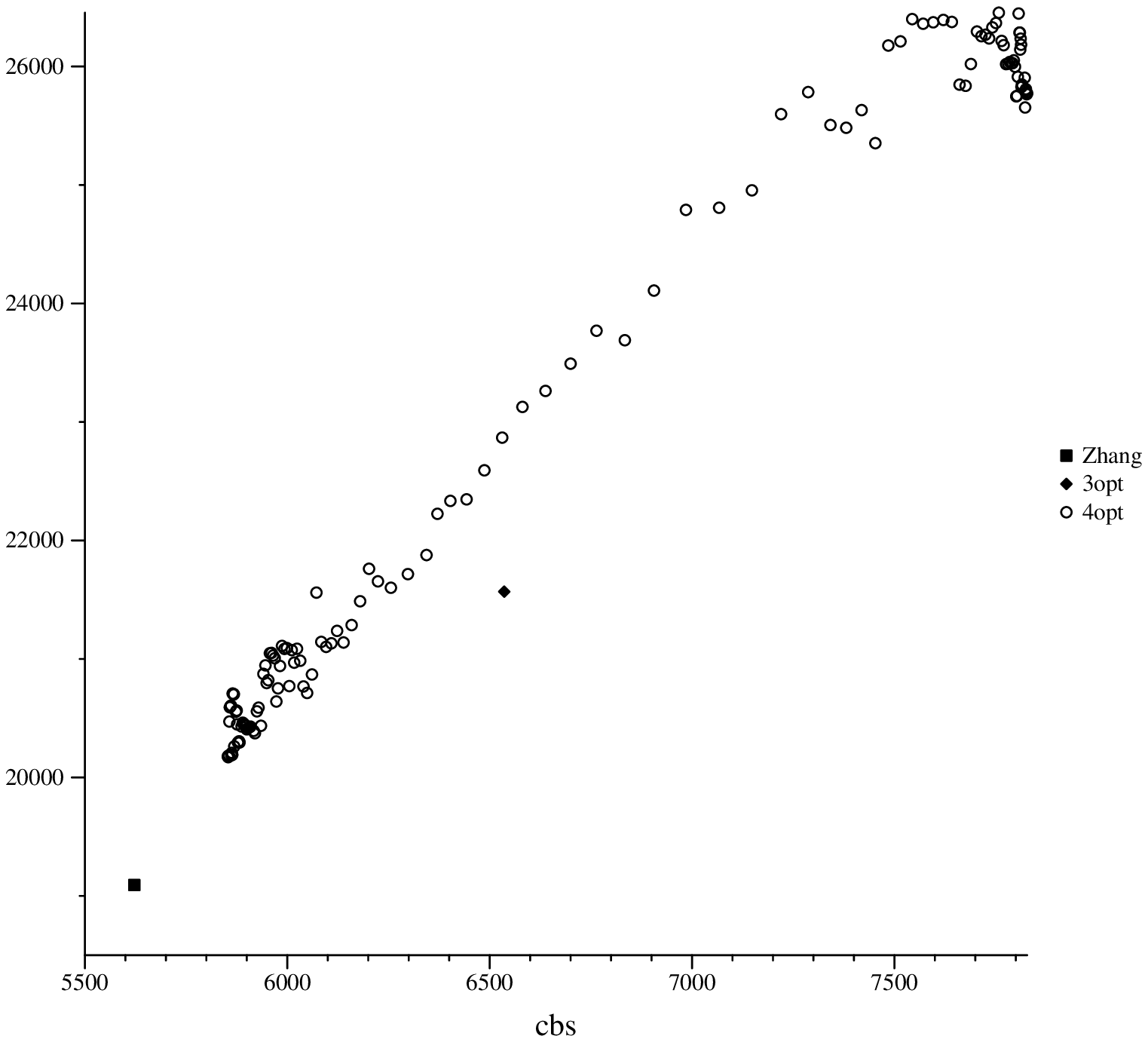}
\figinbox{3in}{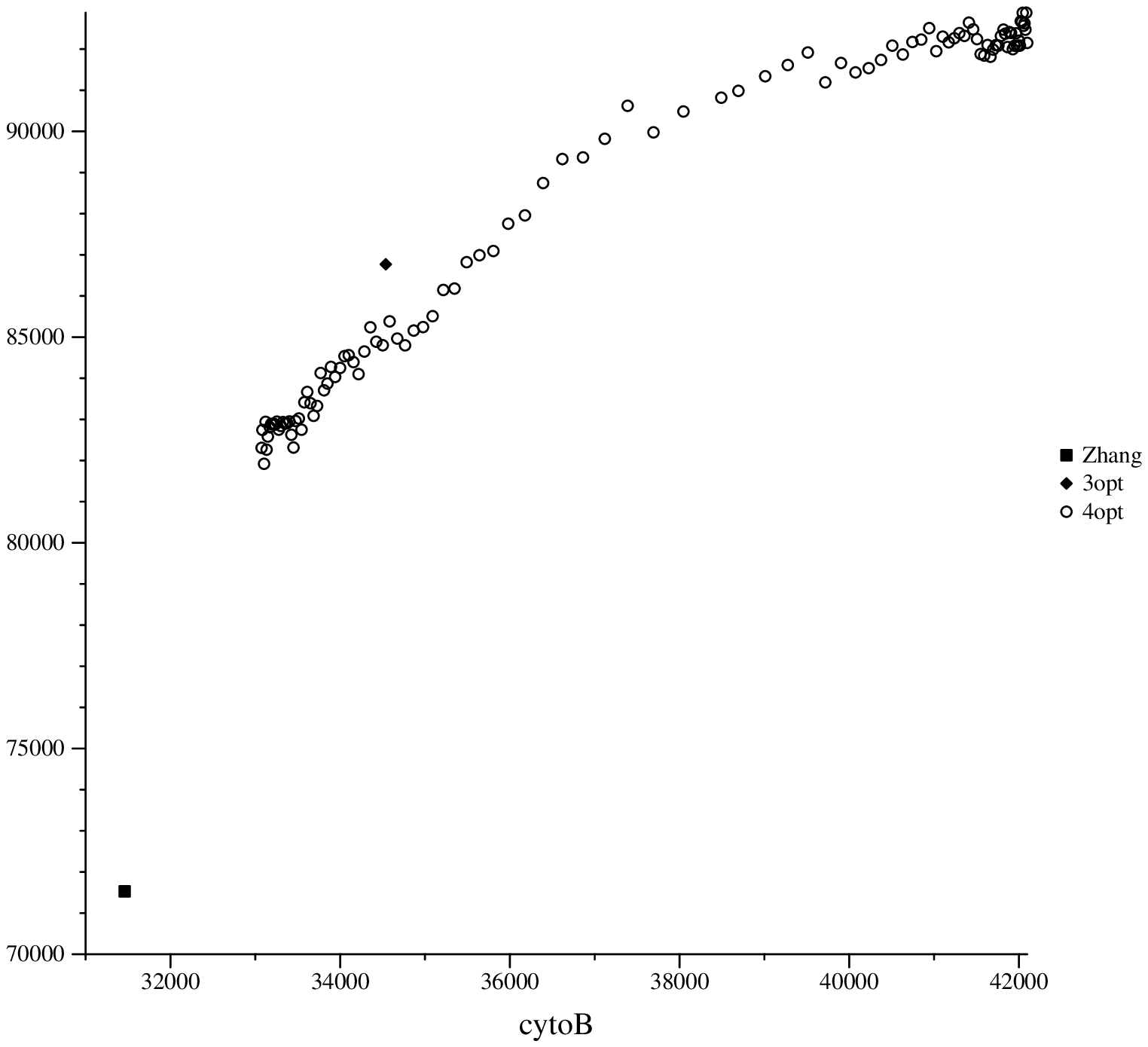}
\caption{Relationship between tour cost (x-axes) and compression
size (y-axes) for {\sc PF00032}, {\sc backPQQ}, {\sc callagen}, {\sc cbs}, and {\sc cytoB},
using the result of Zhang's algorithm,
a 3-opt local optimum,
and a sequence of tours
generated by a series of 4-opt changes.}
\label{fig:tsp2}
\end{figure}

For each file,
we computed various tours on the
corresponding graph $G(\cdot)$.
We computed a close approximation to a TSP
tour using a variation of Zhang's branch-and-bound algorithm \cite{atsp:z93},
discussed by Cirasella et al.~\cite{atsp:c+01}.
We also computed a 3-opt local optimum tour;
and we used a 4-opt heuristic to compute a sequence
of tours of various costs.
Each tour induced an ordering of the columns.
For each column ordering, we computed the optimal, contiguous
partition by DP,
except that
we used GREEDYT on the orderings for {\sc census},
due to computational limitations.
Figures \ref{fig:tsp1} and \ref{fig:tsp2}
plot the results.

The plots demonstrate
a strong, positive correlation
between tour cost and compression performance.
In particular,
each plot shows that the least-cost tour
(produced by Zhang's algorithm)
produced the best compression result.
Table \ref{tab:tspsize}
details the compression improvement
from using the Zhang ordering.
In five files,
Zhang gives an extra compression improvement of at least 5\%
over DP on the original order;
for {\sc cytoB}, the improvement is 20\%.
That the original order for {\sc network}
outperforms the Zhang ordering
is again an artifact of the training/test paradigm.
Figure \ref{fig:tsp1} shows that the
tour-cost/compression-performance
correlation remains strong for this file.

Table \ref{tab:tsphk} displays
the time spent computing Zhang's tour
for each file.
This time is negligible
compared to the time to compute the optimal, contiguous
partition via DP.
(The DP time on {\sc census} is 168531 seconds, four orders of magnitude larger.
For {\sc cytoB}, the DP time is 8640 seconds,
an order of magnitude larger.)
Table \ref{tab:tsphk}
also shows that
Zhang's tour always had cost
close to the Held-Karp lower bound
\cite{hk:hk70,hk:hk71}
on the cost of the
optimum TSP tour.

For off-line training,
therefore,
it seems that computing a good approximation
to the 
TSP reordering
before partitioning
contributes significant compression improvement
at minimal time cost.
Furthermore,
the correlation between tour cost and compression
behaves
similarly to what the theory in Section \ref{sec:tsp} 
would predict
if $\HC(\cdot)$ were sub-additive,
which suggests the existence of some other, similar
structure induced by $\HC(\cdot)$
that would control this relationship.

\begin{table}
\caption{Performance of TSP reordering.
For each, Size is the size of the compressed file
using the Zhang ordering and optimal, contiguous partition
(for {\sc census}, using the GREEDYT partition);
Rate is the corresponding compression rate;
/Gzip is the size relative to gzip;
and /DP is the size relative to
using the optimal, contiguous partition
on the original ordering.}
\label{tab:tspsize}
\begin{center}
\begin{tabular}{l|rr|rr}
 & \multicolumn{2}{c|}{TSP} & & \\
File &
\multicolumn{1}{c}{Size} &
\multicolumn{1}{c|}{Rate} &
\multicolumn{1}{c}{/Gzip} &
\multicolumn{1}{c}{/DP} \\ \hline
{\sc care} & 1199315 & 0.1466 & 0.5890 & 0.9290 \\
{\sc network} & 1822065 & 0.0299 & 0.4859 & 1.0249 \\
{\sc census} & 18113740 & 0.0544 & 0.5901 & 0.8419 \\
{\sc lerg} & 183668 & 0.0528 & 0.4037 & 0.9882 \\
{\sc EGF} & 50027 & 0.0937 & 0.6919 & 0.8843 \\
{\sc LRR} & 48139 & 0.2045 & 0.7796 & 0.9814 \\
{\sc PF00032} & 29625 & 0.0736 & 0.8656 & 0.9685 \\
{\sc backPQQ} & 7131 & 0.3190 & 0.9498 & 0.9923 \\
{\sc callagen} & 51249 & 0.2111 & 0.7611 & 0.8636 \\
{\sc cbs} & 19092 & 0.2586 & 0.8227 & 0.9623 \\
{\sc cytoB} & 71529 & 0.1234 & 0.6522 & 0.7947
\end{tabular}
\end{center}
\end{table}

\begin{table}
\caption{For each file,
the quality of Zhang's tour is expressed as per cent above
the Held-Karp lower bound.
Time is the time in seconds to compute the tour.
}
\label{tab:tsphk}
\begin{center}
\begin{tabular}{l|rr}
File & \multicolumn{1}{c}{\% above HK} & \multicolumn{1}{c}{Time} \\ \hline
{\sc care} & 0.438 & 0.110 \\
{\sc network} & 0.602 & 0.230 \\
{\sc census} & 0.177 & 28.500 \\
{\sc lerg} & 0.011 & 0.010 \\
{\sc EGF} & 0.314 & 0.450 \\
{\sc LRR} & 0.354 & 0.050 \\
{\sc PF00032} & 0.211 & 0.510 \\
{\sc backPQQ}  & 0.196 & 0.050 \\
{\sc callagen} & 0.152 & 0.170 \\
{\sc cbs} & 0.187 & 0.210 \\
{\sc cytoB} & 0.027 & 735.440
\end{tabular}
\end{center}
\end{table}

\section{Complexity of Table Compression: A General Framework}
\label{sec:hardness}

We now introduce a framework 
for studying the computational complexity
of several versions of
table compression problems.
We start with a basic problem
of finding an optimal arrangement of a set of strings
to be compressed.
Given a set of strings,
we wish to compute an order
in which to catenate the strings
into a superstring $X$
so as to minimize the cost of compressing $X$
using a fixed compressor \CC.
To isolate the complexity of finding an optimal order,
we restrict \CC~to prevent it from reordering
the input itself.

Let $x=\sigma_1\cdots\sigma_n$ be a string
over some alphabet $\Sigma$,
and let $\CC(x)$ denote the output of 
\CC~when given input $x$.
We allow \CC~arbitrary time and space,
but we require that it process $x$ monotonically.
That is, it reads the symbols of $x$ in order;
after reading each symbol, it may or may not output
a string.
Let $\CC(x)_j$ be the catenation of all the strings
output, in order, by $\CC$ after processing
$\sigma_1\cdots\sigma_j$.
If \CC~actually outputs a (non-null) string after
reading $\sigma_j$,
then we require that $\CC(x)_j$ must be a prefix of
$\CC(\sigma_1\cdots\sigma_jy)$ for any suffix $y$.
We assume a special end-of-string character not in $\Sigma$
that implicitly terminates every input to $\CC$.

Intuitively, this restriction precludes \CC~from
reordering its input to improve the compression.
Many compression programs used in practice
work within this restriction:
e.g., gzip and compress.

We use $|\CC(x)|$
to abstract the length of $\CC(x)$.
A common measure is bits,
but other measures are more appropriate in certain settings.
For example,
when considering LZ77 compression \cite{comp:zl77},
we will denote by $|\CC(x)|$
the number of phrases in the LZ77 parsing of $x$,
which suffices to capture the length of $\CC(x)$
while ignoring technical details concerning
how phrases are encoded.

Let $X=\{x_1, \ldots, x_n\}$ be a set of strings.  
A {\em batch of} $X$ is an ordered subset of $X$.
A {\em schedule of} $X$ is a
partition of $X$ into batches.
A
batch $B=(x_{i_1}, \ldots, x_{i_s})$ is {\em processed} by $\CC$ by
computing $\CC(B) = \CC(x_{i_1}\cdots x_{i_s})$;
i.e., by compressing the superstring
formed by catenating the strings in $B$ in the order
given.
A schedule $\CS$ of $X$
is {\em processed} by $\CC$ by processing its batches,
one by one,
in any order.
While $\CC(\CS)$ is ambiguous,
$|\CC(\CS)| = \sum_{B\in\CS} |\CC(B)|$ is well defined.
Our main problem can be stated as follows.

\begin{problem}\label{pr:optsched}
Let $X$ be a set of strings.
Find a schedule $\CS$ of $X$ minimizing $|\CC(\CS)|$
among all schedules.
\end{problem}

The classical shortest common superstring (SCS) problem
can be phrased in terms of Problem \ref{pr:optsched}.
For two strings $x$ and $y$,
let $\pref(x,y)$ be the prefix of $x$
that ends at the longest suffix-prefix match of $x$ and $y$.
Let $X$ be a set of $n$ strings,
and let $\pi$ be a permutation of the integers in $[1,n]$.
Define 
$S(X,\pi) = \pref(x_{\pi_1},x_{\pi_2}) \pref(x_{\pi_2},x_{\pi_3})
	\cdots \pref(x_{\pi_{n-1}},x_{\pi_n}) x_{\pi_n}$.
$S(X,\pi)$ is a superstring of $X$;
$\pi$ corresponds to a schedule of $X$;
and the SCS of $X$
is $S(X,\pi)$ for some $\pi$ \cite{strings:g97}.
Therefore, finding the SCS is an instance of Problem \ref{pr:optsched},
where $\CC(\cdot)$ is $S(\cdot)$.
Since finding the SCS is MAX-SNP hard \cite{lass:blty94},
Problem \ref{pr:optsched} is MAX-SNP hard in general.
Different results can hold for specific compressors,
however.

We now formalize table compression problems in this framework.
Consider a table $T$ with $m$ rows and $n$ columns,
each entry a symbol in $\Sigma$. 
Let $T^c$ be the string formed by catenating the columns
of $T$ in order;
let $T^r$ be the string formed by catenating the rows of $T$
in order.

We view $T$ as a set of columns $\{T[1],\ldots,T[n]\}$.
A batch $B=(T[i_1],\ldots,T[i_s])$
then corresponds to a table $T_B = T[i_1]\cdots T[i_s]$,
which we can compress in column- or row-major order.
A column-major order schedule $\CS^c$ of $T$
has compression cost
$|\CS^c| = \sum_{B\in \CS^c} |\CC(T^c_B)|$.
A row-major order schedule $\CS^r$ of $T$
has compression cost
$|\CS^r| = \sum_{B\in \CS^r} |\CC(T^r_B)|$.

\begin{problem}\label{pr:optcol}
Given a table $T$,
find a column-major schedule $\CS^c$ of $T$
minimizing $|\CS^c(T)|$ among all such schedules.
\end{problem}

\begin{problem}\label{pr:optrow}
Given a table $T$,
find a row-major schedule $\CS^r$ of $T$
minimizing $|\CS^r(T)|$ among all such schedules.
\end{problem}

In either column- or row-major order, batches of $T$
are subsets of columns.
In column-major order, each column of $T$ remains a
distinct substring in any schedule.
In row-major order, however, the individual strings
that form a schedule are the row-major renderings
of batches of $T$.
This distinction is subtle yet crucial.
Problem \ref{pr:optcol} becomes equivalent
to Problem \ref{pr:optsched},
so we may consider the latter
in order to establish lower bounds for the former.
Problem \ref{pr:optrow},
however,
is not identical to problem \ref{pr:optsched}:
the row-major order rendering of the batches
results in input strings being intermixed.
We emphasize this distinction
in Section \ref{sec:Runlength},
where we show that,
when \CC~is run length encoding,
Problem \ref{pr:optcol} can be solved in polynomial time,
while Problem \ref{pr:optrow} is MAX-SNP hard.
The connection between table compression and SCS
through Problem \ref{pr:optsched}
makes these problems theoretically elegant
as well as practically motivated.

\section{Complexity with LZ77}
\label{sec:lz77}

We use the standard definitions of L-reduction
and MAX-SNP \cite{oacc:py91}.
Let $A$ and $B$ be two optimization (minimization or maximization)
problems.  
Let $\cost_A(y)$ be the cost of a solution $y$ to some instance of $A$;
let $\opt_A(x)$ be the cost of an optimum solution
for an instance $x$ of $A$;
and define analogous metrics for $B$.
$A$ {\em L-reduces} to $B$ if
there are two polynomial-time functions $f$ and $g$ and constants
$\alpha,\beta>0$ such that:

\begin{itemize}
\item[(1)] Given an instance $a$ of $A$,
$f(a)$ is an
instance of $B$ such that
$\opt_B(f(a)) \leq \alpha\cdot \opt_A(a)$;

\item[(2)] Given a solution $y$ to $f(a)$,
$g(y)$ is a solution to $a$
such that $|\cost_A(g(y))-\opt_A(a)| \leq \beta|\cost_B(y)-\opt_B(f(a))|$.
\end{itemize}

The composition of two L-reductions is also an
L-reduction.
A problem is MAX-SNP hard \cite{oacc:py91} if
every problem in MAX-SNP can be L-reduced to it.
If $A$ L-reduces to $B$,
then if $B$ has a polynomial-time approximation scheme (PTAS),
so does $A$.
A MAX-SNP hard problem is unlikely to have
a PTAS \cite{hap:a+98}.

Now recall the LZ77 parsing rule \cite{comp:zl77},
which is used by compressors
like gzip.
Consider a string $z$, and,
if $|z|\geq 1$, let $z^-$ denote the prefix of $z$
of length $|z|-1$.
If $|z|\geq 2$, then define $z^{--}=(z^-)^-$.

LZ77 parses $z$ into {\em phrases}, each a substring of $z$.
Assume that LZ77 has already
parsed the prefix $z_1\cdots z_{i-1}$ of $z$
into phrases $z_1, \ldots, z_{i-1}$,
and let $z'$ be the remaining suffix of $z$.
LZ77 selects the $i$'th phrase $z_i$ as the
longest prefix of $z'$ that can be obtained by adding a single character to a
substring of $(z_1\cdots z_{i-1}z_i)^-$.
Therefore, $z_i$ has the
property that $z_i^-$ is a substring of $(z_1z_2\cdots
z_{i-1}z_i)^{--}$, but $z_i$ is not a substring of $(z_1z_2\cdots
z_{i-1}z_i)^{-}$.
This recursive definition is sound \cite{lec:km00}.

After parsing $z_i$, LZ77 outputs an
encoding of the triplet $(p_i, \ell_i, \alpha_i)$, where $p_i$ is the
starting position of $z_i^-$ in $z_1z_2\cdots z_{i-1}$;
$\ell_i=|z_i|-1$; and $\alpha_i$ is the last character of $z_i$.
The length of the encoding is linear in the number of phases,
so
when \CC~is LZ77, we denote by $|\CC(z)|$ the number of phrases
in the parsing of $z$.
This cost function
is commonly used to establish the performance of LZ77
parsing \cite{it:ct91,lec:km00}.

\subsection{Problem \ref{pr:optsched} }\label{sec:LZ77hard}

We show that Problem \ref{pr:optsched}
is MAX-SNP hard
when \CC~is LZ77.
Consider TSP(1,2), the traveling salesman problem
on a complete graph
where each distance is either 1 or 2.
An instance of TSP(1,2) can be specified by a
graph $H$, 
where the edges of $H$ connect those pairs of vertices with distance 1.
The problem remains MAX-SNP hard
if we further restrict the problem so that the degree of each vertex
in $H$ is bounded by some arbitrary but fixed constant \cite{tsp:py93}.
This result holds for both symmetric and asymmetric TSP(1,2);
i.e., for both undirected and directed graphs $H$.
We assume that $H$
is directed. 
The following lemma shows that we may also assume
that no vertex in $H$
has outdegree 1.

\begin{lemma}
TSP(1,2) L-reduces to TSP(1,2) with the additional stipulation
that no vertex has only one outgoing cost-1 edge.
\end{lemma}
\begin{proof}
Consider instance $A$ of TSP(1,2).
For each vertex $v$ with only one outgoing cost-1 edge,
to some $v'$,
we create a new vertex $v''$
such that edges $(v,v'')$, $(v'',v)$, and $(v'',v')$
have cost 1
and all other edges incident on $v''$ have cost 2.
Thus we form instance $B$.
A solution to $B$ is mapped to a solution to $A$
by splicing out all newly created vertices.
If $A$ has $n$ vertices,
then $B$ has at most $2n$ vertices.
All solutions to both have cost $O(n)$,
so we need only prove that the reverse mapping of solutions
preserves optimality.

Assume $S_B$ is an optimal solution to $B$ and $S_A$ the corresponding
mapped solution to $A$.
Note that $\cost(S_A) \leq \cost(S_B) - \eta$,
where $\eta$ is the number of vertices
created to form $B$.
(We drop the subscripts to $\cost(\cdot)$,
as there is no ambiguity.)
If $S_A$ is not optimal,
there is some $S'_A$ such that $\cost(S'_A) < \cost(S_A)$.
We can form a solution $S'_B$ to $B$
by replacing each edge $(v,z)$ in $S'_A$,
where $v$ has only one cost-1 outgoing edge,
with edges $(v,v')$ and $(v',z)$.
This gives $\cost(S'_B) = \cost(S'_A) + \eta < \cost(S_A) + \eta \leq \cost(S_B)$,
contradicting the optimality of $S_B$.
\end{proof}

We associate a set $S(H)$ of strings to the vertices and edges of $H$;
$S(H)$ will be the input to Problem \ref{pr:optsched}.
Each vertex $v$ engenders three symbols:  $v$, $v'$, and $\$_v$.
Let
$w_0, \ldots, w_{d-1}$ be the vertices on the edges out of $v$ in $H$,
in some arbitrary but fixed cyclic order.
For $0\leq i< d$ and mod-$d$ arithmetic, we say that
edge $(v,w_i)$ {\em cyclicly precedes} edge $(v,w_{i+1})$.
The $d+1$ strings we
associate to $v$ and these edges are:
$e(v,w_i)=(v'w_{i-1})^4v'w_{i}$,
for $0 \leq i<d$
and mod-$d$ arithmetic;
and
$s(v)=v^4(v')^5\$_v$. 
That $d\not=1$ implies that $w_i\not=w_{i+1}$
when $d\not=0$,

To prove MAX-SNP hardness,
we first
show how to transform a TSP(1,2) solution for $H$ into a solution
to Problem \ref{pr:optsched} with input $S(H)$.
We then show how to transform in polynomial time a solution to Problem
\ref{pr:optsched} into a TSP(1,2) solution of a certain cost.
We use the intermediate step
of transforming the first solution into a canonical form
of at most the same cost.

The canonical form solution will correspond to the required
TSP(1,2) tour.
We will show that, for all edges $(v,w)$,
$e(v,w)$ will parse into one phrase when
immediately preceded by $e(v,y)$
for the edge $(v,y)$ that cyclicly precedes $(v,w)$,
and into more than one phrase otherwise;
and we will show that $s(v)$ will parse into two phrases
when immediately preceded by $e(x,v)$ for some edge $(x,v)$,
and into three phrases otherwise.
Thus,
an edge $(v,w_i)$ in the path will best be encoded
as 
$s(v)e(v,w_{i+1})e(v,w_{i+2})\cdots e(v,w_{i}) s(w_i)$.
This is the core idea of our canonical form.

Lemmas \ref{lem:boundary}--\ref{lem:parsecount}
provide a few needed facts about the parsing of strings in $S(H)$.
In what
follows, $X$ denotes both a batch in $S(H)$ and the string obtained by
catenating the strings in the batch in order.

\begin{lemma}\label{lem:boundary}
Let $X=x_1\cdots x_s$ be a batch of $S(H)$,
where each $x_i$ is $s(v)$ for some vertex $v$ or $e(v,w)$
for some edge $(v,w)$. 
For each $1 \leq j
\leq s$, some phrase in the LZ77 parsing of $X$
ends at the last symbol of $x_j$.
\end{lemma} 
\begin{proof}
The proof is by induction. 
The base case is for $j=1$. 
If $x_1=s(v)$ for some vertex $v$,
then the lemma holds, because $\$_v$ 
appears only at the end of $s(v)$.
Otherwise,
$x_1=e(v,w_i)$ for some edge $(v, w_i)$.
Since $x_1$ appears first in $X$, its parsing is $v'$, $w_{i-1}$,
$(v'w_{i-1})^3v'w_{i}$.
(That no vertex has outdegree one
implies that $w_i\not=w_{i-1}$.)
The lemma again holds.

Assume by induction that the
lemma is true up through the parsing of $x_{j-1}$;
we show that it
holds for the parsing through $x_j$.
Again, if $x_j=s(v)$ for some vertex $v$, the lemma is true,
because $\$_v$
appears only at the end of $s(v)$.
Otherwise, $x_j=e(v,w_i)$, for
some edge $(v, w_i)$.
There are two cases.
\begin{enumerate}

\item
$x_{j-1}=e(v,w_{i-1})$. 
Then
$x_{j-1}x_{j}=(v'w_{i-2})^4(v'w_{i-1})^5v'w_{i}$.
By induction,
a phrase ends at the first occurrence of $w_{i-1}$.
Thus, the next phrase is $(v'w_{i-1})^4v'w_{i}=x_j$.

\item
$x_{j-1} \neq e(v,w_{i-1})$.
Again by induction, the first phrase, say $c$,
of the parsing that overlaps
$x_j$ must start at the first character of $x_j$.
Since
$(v'w_{i-1})^2$ does not occur in $x_1\cdots x_{j-1}$,
the first phrase cannot
extend past the fourth character of $x_j$.
We have the following subcases.

\begin{enumerate}

\item
\label{it:A}
$c$ ends at the first character of $x_j$.
Therefore $v'$
does not occur in $x_1 \cdots x_{j-1}$. Since
$x_j=(v'w_{i-1})^4v'w_{i}$, we have that the phrase following $c$, say
$c'$, must be either $w_{i-1}$ or $w_{i-1}v'$,
depending on whether or not
$w_{i-1}$ occurs in $x_1 \cdots x_{j-1}$.
(1) When $c'$ is $w_{i-1}$, the
next phrase is $(v'w_{i-1})^3v'w_{i}$ and ends on the last character 
of $x_j$, as required.
(2) When $c'$ is $w_{i-1}v'$, the next phrase is
$(w_{i-1}v')^3w_{i}$, again completing the induction.

\item
Remaining cases.
When $c$ ends at the second and third
character of $x_j$, the result follows as in (\ref{it:A}.1) and (\ref{it:A}.2),
respectively.
When $c$ ends at the forth character, the next
phrase is $(v'w_{i-1})^2v'w_{i}$ and ends at the last character of $x_j$
as required.
\end{enumerate}
\end{enumerate}
\end{proof}

\begin{lemma}\label{lem:vertices}
Let $X$ be a batch of $S(H)$
and $v$ be any vertex such that $s(v)\in X$.
If $s(v)$ is immediately preceded by $e(q,v)$ for some 
edge $(q,v)$,
$s(v)$ is parsed into precisely two phrases
during the parsing of $X$;
otherwise,
$s(v)$ is parsed into precisely three phrases.
\end{lemma}
\begin{proof}
Assume first that $s(v)$ is immediately preceded by $e(q,v)$ for some 
edge $(q,v)$.
Then
$e(q,v) s(v)=(q'z)^4q'vv^4(v')^5\$_v$
for some $z$. By Lemma \ref{lem:boundary},
a phrase of the parsing must end with the last character of $e(q,v)$.  
Since $v^4$ does not appear elsewhere in $X$, 
the next two phrases of the parsing must be $v^4v'$ and $(v')^4\$_v$.

In the other case,
$v^2$ does not occur to the left of 
$s(v)$ in $X$.
Again using Lemma \ref{lem:boundary}, the parsing of
$X$ has a phrase starting
at $s(v)$.
If $v$ appears
to the left of $s(v)$ in $X$, the parsing produces
$v^2$, $v^2v'$, and $(v')^4\$_v$;
Otherwise, it produces $v$, $v^3v'$, and $(v')^4\$_v$.
\end{proof}

\begin{lemma}\label{lem:edges}
Let $X$ be a batch of $S(H)$ and $(v,w)$ be any edge
such that $e(v,w)\in X$.
Let $(v,y)$ be the edge that cyclicly precedes $(v,w)$.
If $e(v,w)$ is immediately preceded in $X$ by $e(v,y)$,
then $e(v,w)$ is parsed into precisely one phrase
during the parsing of $X$;
if $e(v,w)$ is immediately preceded by $s(v)$,
then $e(v,w)$ is parsed into precisely two phrases;
in any other case,
$e(v,w)$ is parsed into at least two phrases.
\end{lemma}
\begin{proof}
By Lemma~\ref{lem:boundary}, some phrase
starts at the first character of $e(v,w)$. 
Assume $e(v,y)$ immediately precedes $e(v,w)$;
$e(v,y)e(v,w) = (v'z)^4v'y (v'y)^4v'w$ for some $z$.
The parsing of $e(v,w)$ produces
the one phrase $(v'y)^4v'w=e(v,w)$.
(Nowhere else does this string appear in $X$.)

Assume $s(v)$ immediately precedes $e(v,w)$;
$s(v)e(v,w) = v^4(v')^5\$_v (v'y)^4v'w$.
If $v'y$ occurs earlier in $X$,
the parsing of $e(v,w)$ produces
phrases $v'yv'$ and $(yv')^3 w$,
because $v'yv'$ cannot occur elsewhere.
Otherwise,
the parsing produces $v'y$ and $(v'y)^3 v'w$.

In any other case, $e(v,w)$ is preceded by a character
other than $v'$.
If $v'$ occurs earlier in $X$,
then the parsing of $e(v,w)$ produces two phrases
as in the case of $s(v)$ preceding $e(v,w)$.
Otherwise, the parsing produces
$v'$ and then at least one more phrase.
\end{proof}

Now define a schedule $Y_1, \ldots, Y_t$ to be
{\em standard} if and only if:
for each batch $Y_i$,
the order in which the strings $s(v)$ appear in $Y_i$
corresponds to a path in $H$;
the paths
associated to $Y_i$ and $Y_j$ are disjoint
for each $i\not= j$;
and each vertex of $H$ appears as $s(v)$
in some batch $Y_i$.

We give a polynomial time algorithm that transforms a
schedule $\CS = (X_1, \ldots, X_g)$ into
a standard schedule that parses into no more phrases than does \CS.
The algorithm consists of two phases.
The first phase
computes a set of disjoint paths that covers
all the vertices of $H$.
It iteratively combines paths, guided by \CS,
until no further combination is possible.
The second phase
transforms each path into a batch
such that the resulting schedule is standard.

\centerline{\bf Algorithm STANDARD}

\begin{itemize}

\item[{\bf P1}] 
\begin{enumerate}
\item
Place each vertex $v$ of $H$ in a single-vertex path. 
If $s(v)$ 
is the first string in some batch
in \CS,
label $v$ terminal;
otherwise, label $v$ nonterminal.

\item
While there exists a path with nonterminal left end
point, pick one such end point $v$ and process it as follows.
Let $X_i$
be the batch in which $s(v)$ occurs.
Let $x(u)$ be the string
(associated to either vertex $u$ or to one of its outgoing edges)
that precedes $s(v)$ in $X_i$. If $x(u)$ ends in a symbol other than
$v$, label vertex $v$ terminal. Otherwise, $(u,v)\in H$,
so connect $u$ to $v$,
and, for each edge $(u, w) \in H$, $u\not=w$, such that
$s(w)$ is immediately preceded by $e(u,w)$, declare $w$ terminal.
(This guarantees that Phase One actually builds paths.)
\end{enumerate}

\item[{\bf P2}] Let $\CA_1, \ldots, \CA_t$ be the paths
obtained at the end of Phase One. 
We transform each path $\CA_j$ into a batch $Y_j$.
If $\CA_j$ consists of a single vertex $v$,
then $Y_j$
consists of $s(v)$ followed by all the $e(v,w_j)$'s arranged in
cyclic order.

Otherwise, $\CA_j$ contains more than one vertex.
Initially $Y_j$ is empty.
For each edge $(u,v)$ in order in the path,
we append to $Y_j$:
$s(u)$ followed by all of its $e(u,w_j)$'s,
in cyclic order ending with $e(u,v)$.
When there are no more edges to process, the last
vertex of the path is processed as in the singleton-vertex case.
\end{itemize}

\begin{lemma}
\label{lem:standard}
In polynomial time,
Algorithm STANDARD transforms
schedule $X_1, \ldots, X_g$
into a standard schedule $Y_1, \ldots,Y_t$ of no higher cost.
\end{lemma}
\begin{proof}
That Algorithm STANDARD runs in polynomial
time and $Y_1,\ldots,Y_t$ is standard
follow immediately from the specification.

We now show that each batch $Y_j$
parses into no more phrases
than do its corresponding components 
in the input schedule.
Consider the path,
$(v_1, v_2, \ldots, v_r)$ from which $Y_j$
is derived.
Let $d(v)$ be
the outdegree of any vertex $v$.
$Y_j = s(v_1) e(v_1,w^1_1) \cdots e(v_1,w^1_{d(v_1)})
	\cdots
	s(v_r) e(v_r,w^r_1) \cdots e(v_r,w^r_{d(v_r)})$,
where the $w^i_j$'s are the neighbors in cyclic order
out of $v_i$
and,
for $1\leq i < r$,
we assume without loss of generality that $w^i_{d(v_i)} = v_{i+1}$.

By Lemma \ref{lem:vertices},
for $2\leq i\leq r$,
$s(v_i)$ parses into two phrases,
which is optimal.
By Lemma \ref{lem:edges},
for $1\leq i\leq r$ and $2\leq j \leq d(v_i)$,
$e(v_i,w^i_j)$ parses into one phrase,
which is optimal.
We thus need only consider the parsing
of $s(v_1)$ and, for $1\leq i\leq r$, $e(v_i,w^i_1)$.

The strings $e(v_i,w^i_1)$, for $1\leq i\leq r$ each parse into two 
phrases in $Y_j$, by Lemma \ref{lem:edges}.
There must be some 
$e(v_i,x)$ that is not immediately preceded by its
cyclic predecessor in some $X_k$,
and this instance of $e(v_i,x)$ also parses into at least two phrases,
by Lemma \ref{lem:edges}.
This accounts for the first $e(\cdot)$ string immediately
following each $s(\cdot)$ string in $Y_j$.

Finally, if $s(v_1)$ is not immediately preceded by some $e(v,v_1)$
in the input batch $X_k$ in which $s(v_1)$ appears,
we are done,
for $s(v_1)$ is parsed into three phrases in both $X_k$ and $Y_j$,
by Lemma \ref{lem:vertices}.
Otherwise, 
consider the maximal sequence
$e(v,w_a) e(v,w_{a+1}) \cdots e(v,w_{a+\ell}=v_1)s(v_1)$
in $X_k$,
where the $w_a$'s are cyclicly ordered neighbors of $v$.
Because STANDARD declared $v_1$ to be terminal,
there was another edge $(v,y)$ such that $e(v,y)$
immediately preceded $s(y)$ in some $X_{k'}$,
which STANDARD used to connect $v$ and $y$ in some path.
This engenders an analogous maximal chain of
$e(v,\cdot)$ strings followed by $s(y)$ in $X_{k'}$.

Thus, 
there are at least two strings $e(v,\cdot)$ not immediately
preceded in the input by their cyclic predecessors;
Lemma \ref{lem:edges} implies each
is parsed into at least two phrases.
We can charge the extra phrase generated by $s(v_1)$ in $Y_j$
against one of them,
leaving the other
for the extra phrase in the parsing of the 
$e(v,\cdot)$ phrase immediately following $s(v)$ in some $Y_{j'}$.
\end{proof}

\begin{lemma}
\label{lem:parsecount}
A batch $Y_j$ output by STANDARD,
corresponding to a path $(v_1,\ldots,v_r)$,
parses into $3r + 1 + \sum_{i=1}^r d(v_i)$ phrases.
\end{lemma}
\begin{proof}
By Lemma \ref{lem:vertices},
each $s(\cdot)$ string parses into 2 phrases,
except $s(v_1)$, which parses into 3,
contributing $2r+1$ phrases.
Lemma \ref{lem:edges}
implies
that each $e(\cdot)$ parses into 1 phrase,
except each following an $s(\cdot)$,
which parses into 2,
contributing $r+\sum_{i=1}^r d(v_i)$ phrases.
\end{proof}

\begin{theorem}
Problem \ref{pr:optsched} is MAX-SNP hard when $\CC$ is LZ77.
\end{theorem}
\begin{proof}
Let the graph $H$ defined at the beginning of the section have
$n_h$ vertices and $m_h$ edges. 
Let $k$ be the minimum number of cost-2 edges that
suffice to form a TSP(1,2) solution.
Then the cost of the solution
is $n_h-1+k$. 
Associating strings to vertices and
edges of $H$, as discussed above, 
we argue that the optimal
schedule for those strings produces $m_h+k+3n_h+1$ phrases.  
The reduction is linear, since $m_h =O( n_h)$
by the assumption of bounded outdegree.

Assume that the TSP(1,2) solution
with $k$ cost-2 edges is the path
$v_1, v_2, \ldots, v_{n_h}$. 
Then in polynomial time we can construct a
corresponding standard schedule of the form output by STANDARD,
which Lemma \ref{lem:parsecount}
shows parses into $m_h+k+3n_h+1$ phrases.

For the converse, assume that we are given a schedule of cost
$m_h+k+3n_h+1$.  
By Lemma \ref{lem:standard},
we can transform it
in polynomial time into a standard schedule $Y_1, \ldots, Y_t$ of no
higher cost.
Recall that to each batch we can associate a path of
$H$.
Let $v_1,v_2, \ldots, v_{n_h}$ the an ordering of the vertices of $H$
corresponding to an arbitrarily chosen processing order for the
sequence of batches.
Then, $H$ cannot be missing more than $k$ of the
edges $(v_i, v_{i+1})$,
or else, by Lemma \ref{lem:parsecount},
the cost of the standard schedule would
exceed $m_h+k+3n_h+1$.
\end{proof}

\section{Complexity with Run Length Encoding}\label{sec:Runlength}
\label{sec:rle}

In run length encoding (RLE),
an input string is parsed into phrases of the form
$(\sigma,n)$, 
where $\sigma$ is a character, and $n$ is the number of times
$\sigma$ appears consecutively.
For example, $aaaabbbbaaaa$ is parsed into $(a,4)(b,4)(a,4)$.

\subsection{Problem \ref{pr:optsched}}\label{sec:Runstringhard}

\begin{theorem}
Problem \ref{pr:optsched} can be solved in polynomial time
when $\CC$ is run length encoding.
\end{theorem}
\begin{proof}
Let $x_1,\ldots,x_n$
be the input strings.
We can assume without loss of generality that each $x_i$
is of the form $\sigma\sigma'$;
i.e., two distinct characters.
The parsing of any characters between them cannot be optimized
by rearranging the strings.
Furthermore, if $x_i=\sigma\sigma$,
we can simply merge $x_i$ with another string, $x_j$,
that begins or ends with $\sigma$;
if no such $x_j$ exists,
we can ignore $x_i$ completely,
since again its parsing cannot be optimized
by rearrangement.

We claim that a shortest common superstring (SCS) of the input
corresponds to an optimal schedule.
As described earlier,
an SCS is $\pref(x_{\pi_1},x_{\pi_2})\cdots\pref(x_{\pi_{n-1}},x_{\pi_n})x_{\pi_n}$ for some permutation $\pi$.
Note that $\pref(x_i,x_j)$ is of length 2 if the last character
of $x_i$ equals the first of $x_j$ and 3 otherwise.
Thus, an SCS gives an optimal RLE parsing,
and SCS can be solved in polynomial time
when all input strings are of length two \cite{gandj:gj}.
\end{proof}

\subsection{Problem \ref{pr:optrow}}\label{sec:RunLengthT}

As in Section \ref{sec:LZ77hard}, we transform
the vertices and edges of $H$ into an instance of Problem
\ref{pr:optrow}.
We associate a column to each vertex and edge of $H$.

For each vertex $v$,
we generate three symbols: $v$, $v'$, and $v''$.
Let $w_0, \ldots, w_{d-1}$ be the
vertices on the edges out of $v$ in some fixed, arbitrary cyclic order.
We associate the following strings to $v$ and its outgoing edges:
$s(v)=v'v''v$;
and $e(v,w_i)=v'v''w_i$, $0 \leq i<d$.
The input table is formed by assigning each such string,
over all the vertices,
to a column.

Consider a TSP(1,2) solution
with $k$ cost-2 edges.
We can arrange the induced strings
into a table $T$ describing these paths.
Place all strings
corresponding to a vertex $v$ in a contiguous interval of the table with
$s(v)$ being the first column of the interval.
For any edge $(v,q)$ 
in the collection
of paths, place
the interval corresponding to $q$
immediately after that corresponding to $v$,
and place the string $e(v,q)$
last in the interval for $v$;
otherwise,
the order of the intervals
and of the strings corresponding to edges
can be arbitrary.
We say the table is in {\em standard form}
for the collection of paths.

\begin{theorem}
Problem \ref{pr:optrow}
is MAX-SNP hard for tables of at least 3 rows
when $\CC$ is run length encoding.
\end{theorem}
\begin{proof}
We prove the theorem for three rows first and then extend it to
larger numbers of rows.
Let $n_h$ and $m_h$ be the number
of vertices and edges in $H$, rsp.,
and let $n=n_h+m_h$ be the number of columns
in the induced table.
Associate strings to the vertices and edges
as described above.  
Let $k$ be the minimum number cost-2
edges that suffice
to form a TSP(1,2) solution for $H$.
Then the cost of
the solution is $n_h-1+k$.
Let $v_1,v_2, \ldots,v_{n_h}$ be an ordering of the vertices
in $H$ corresponding to the $k+1$ disjoint paths.
Let $T$ be the corresponding standard form table.
Let $S$ be the schedule obtained by taking as a batch each interval
of the table corresponding to a path.
The row-major cost of $S$ is
$2n+m_h+k+1$. This  completes one direction of the transformation.

As for the other direction, assume that we are given a solution to the
instance of optimum table compression that has cost $2n+m_h+k+1$.
Let
$T'$ be the table of the solution schedule.
In polynomial time,
we can transform $T'$ into a standard form table $T$
with a schedule of at most the same cost.
We simply observe that, if the $e(\cdot)$
and $s(\cdot)$ strings for any vertex
are not contiguous,
we can rearrange the columns to make them so,
saving at least two phrases and generating at most two
in the new parsing.

Since a table in standard form
corresponds to an ordering of the vertices,
it must be that $H$ cannot be missing more than $k$ edges,
or else
the cost of the table in standard form would be greater than 
$2n+m_h+k+1$.

When the number of rows $m$ exceeds three,
we
use one additional character $\$$.
Each string is as in the case $m=3$, except that now
is augmented to end with the suffix $\$^{m-3}$. This would add one more
phrase to the parsing of the set of strings, and the linearity of the
transformation still holds.
\end{proof}

\section{Conclusion}
\label{sec:conc}

We demonstrate a general framework
that links independence among groups of variables
to efficient partitioning algorithms.
We provide general solutions in ideal cases
in which dependencies form equivalence classes
or cost functions are sub-additive.
The application to table compression
suggests that 
there also 
exist weaker structures 
that
allow partitioning
to produce significant cost improvements.
Open is the problem of refining the theory
to explain these structures
and extending it to other applications.

Based on experimental results,
we conjecture that our TSP reordering algorithm
is close to optimal;
i.e., that no partition-based algorithm
will produce significantly better compression rates.
It is open if there exists a measurable lower bound
for compression optimality,
analogous, e.g., to the Held-Karp TSP lower bound.

Finally, while we have shown some MAX-SNP hardness results
pertaining to table compression,
it is open whether the problem is even approximable
to within constant factors.

\section*{Acknowledgements}
We are indebted to David Johnson
for running his implementation of Zhang's algorithm
and local 3-opt
on our files.
We thank
David  Applegate,
Flip Korn,
Cecilia LaNave,
S.~Muthukrishnan,
Grazieno Pesole,
and
Andrea Sgarro
for many useful discussions.

\bibliographystyle{plain}
\bibliography{align,approx,complexity,compress,facts,infth,optimization}

\end{document}